\documentclass[aps,pra,superscriptaddress,twocolumn,reprint]{revtex4-1} 
\usepackage{bm} \usepackage{graphicx} \usepackage{amsmath}
\usepackage{amsthm}
\usepackage{amssymb} \usepackage{txfonts}
\usepackage{subfigure}

\DeclareRobustCommand\openzero{\leavevmode\hbox{0\kern-.55em0}}
\mathchardef\minus="002D

\newcommand{\ket}[1]{|{#1}\rangle}
\newcommand{\bra}[1]{\langle{#1}|}
\newcommand{\mean}[1]{\langle{#1}\rangle}

\newcommand\afm{\ket{\rm AFM}}
\newcommand\fm{\ket{\rm FM}}
\newcommand\ua\uparrow
\newcommand\da\downarrow

\begin{document}

\title{Optimal Quench for Distance-Independent Entanglement
and Maximal Block Entropy}

\author{Bedoor Alkurtass}
\affiliation{Department of Physics and Astronomy, University College London, Gower Street, WC1E 6BT London, United Kingdom}
\affiliation{Department of Physics and Astronomy, King Saud University, Riyadh 11451, Saudi Arabia}

\author{Leonardo Banchi}
\affiliation{Department of Physics and Astronomy, University College London, Gower Street, WC1E 6BT London, United Kingdom}

\author{Sougato Bose}
\affiliation{Department of Physics and Astronomy, University College London, Gower Street, WC1E 6BT London, United Kingdom}

\begin{abstract}
We optimize a quantum walk of multiple fermions
following a quench in a spin chain to generate near ideal resources for quantum networking. 
We first prove an useful theorem mapping the correlations evolved from specific quenches to the apparently unrelated problem of quantum state transfer between distinct spins.   
This mapping is then exploited 
to optimize the dynamics and produce large amounts of entanglement distributed in very special ways. 
Two applications are considered: the simultaneous generation of 
many Bell states between pairs of distant spins (maximal block entropy),
or high entanglement between the ends of an arbitrarily long chain
(distance-independent entanglement).
Thanks to the generality of the result, we study its implementation 
in different experimental setups using present technology: NMR, ion traps and ultracold atoms in optical lattices. 
\end{abstract}

\date{\today}

\maketitle

\section{Introduction}
Entanglement is an essential resource for linking distinct quantum registers
through teleportation 
\cite{bennett_teleporting_1993}.
Therefore, the generation and distribution of entanglement over 
distances long enough to link separated quantum units on a chip
is becoming topical
\cite{moehring_entanglement_2007,
  hucul_modular_2014,weitenberg_quantum_2011,yao_robust_2011,banchi_nonperturbative_2011, ping_practicality_2013}.
In parallel, the spread/growth of entanglement from unentangled states due to a
quench has become a topic of great interest in condensed matter, both
theoretically
\cite{calabrese_evolution_2005,de_chiara_entanglement_2006,das_entanglement_2011,barmettler_quantum_2008,sengupta_entanglement_2009,sadiek_entanglement_2010,alkurtass_entanglement_2011,sadiek_dynamics_2013,eisert_towards_2004,schachenmayer_entanglement_2013,hauke_spread_2013,kim_ballistic_2013,bardarson_unbounded_2012,serbyn_universal_2013,vosk_many-body_2013} 
and experimentally
\cite{jurcevic_observation_2014,richerme_non-local_2014}. Could it be used for the practical purpose of connecting quantum registers, especially in view of its experimental viability \cite{fukuhara_microscopic_2013,jurcevic_observation_2014,richerme_non-local_2014}? 
This is largely unexplored as the entanglement generated in typical quenches is between 
blocks and is {\it not} arranged in the special form of a high
entanglement between individual spin pairs. Such block-block entanglement is
not readily useful for applications such as connecting quantum registers. 
Moreover, whether two large complementary parts of a spin chain can get
{\it maximally} entangled through a
quench with no further manipulation/control is not known to date.
 Here we show that a simple global quench of a Hamiltonian with spatially varying couplings can not only yield maximal entanglement between complementary blocks of a spin chain, but this entanglement is also distributed in the special form of singlet states of individual pairs of spins. This state is a resource for many simultaneous entangling gates between pairs of quantum registers. 
 From a condensed matter perspective, we show that {\it unbounded} entanglement (a topic of high interest \cite{bardarson_unbounded_2012})
 can be obtained even in non-interacting systems 
 with suitably engineered interactions.

Entanglement between extremal spins of a chain is currently a
topic of active interest because of its linking power
\cite{campos_venuti_long-distance_2006,campos_venuti_long-distance_2007,giampaolo_long-distance_2010,zippilli_adiabatic_2013,wichterich_exploiting_2009,alkurtass_quench_2013,bayat_entanglement_2010,sodano_kondo_2010,bayat_initializing_2011}.
However, whether fast (non-adiabatic) dynamics following a ``global" quench
can produce {\it distance-independent} entanglement between extremal spins of
a chain is an open question. In schemes studied so far, it typically falls off
with the length of the chain
\cite{wichterich_exploiting_2009,alkurtass_quench_2013,bayat_entanglement_2010,sodano_kondo_2010}.
This task is, of course, less demanding than maximal entanglement between two complementary blocks. As a second result, we show that distance independent entanglement can be achieved with quenches to Hamiltonians with ``minimal" spatial variation of couplings. 

 Our study is also motivated by some recent experimental activities. Multiparticle quantum
walks have recently gathered exceptional interest
\cite{sansoni_two-particle_2012}  fueled largely by the inefficiency of the 
classical simulation of many-boson quantum walks \cite{aaronson_computational_2011}. The spin chain we consider can be mapped into a fermion
hopping model, so the quench generates a 
many-fermion quantum walk. Many-fermion quantum walks, on the other hand, can be efficiently simulated classically
\cite{terhal_classical_2002} -- so one might naively assume them to be of no advantage to quantum information processing. In that respect, what we find here is interesting -- the many fermion quantum walk can still generate resources for quantum information tasks, 
and produce a maximal amount of entanglement. Another subject of high
experimental interest is 
the spread of quantum correlations
 after a quench
 \cite{cheneau_light-cone-like_2012,jurcevic_observation_2014,richerme_non-local_2014}.
 Our work is aimed at offering possibilities to take these studies beyond their
 fundamental remit,  namely in ``optimizing'' these correlations for useful purposes.


In this article we consider the quench that evolves an initial N\'eel or
ferromagnetic state according to an appropriate free fermion spin model in 1D
with the possibility of a spatial variation of the spin-spin couplings. First we prove a mapping which
relates the resulting many-fermion quantum walk to a simpler dynamics:
we show that correlations developing between two sites $m$,$n$ at a time $t/2$  after the quench
is related to amplitude of single walker to travel from $m$ to $n$ in a time
$t$. The difference in the times is noteworthy, and highlights the
non-triviality of the result (we are {\em not} merely restating the well known
resolvability of the many free fermion walk to simultaneous single particle
walks). 
Such continuous time quantum walks of a single walker has been
subject to intensive research in recent years, motivated by the understanding
of quantum information transfer 
\cite{bose_quantum_2003} and quantum computation \cite{childs_universal_2009}. 
Our new mapping between quench and quantum walk, together with the
wealth of results about information transmission, allows us to propose 
different optimal strategies to dynamically 
generate entanglement between distant sites, simultaneously.
In particular, exploiting Hamiltonians which allow
perfect transmission \cite{kay_perfect_2010},
one can generate with the quench a maximal set of Bell states between
distant spins and ultimately maximal block entropy. 
A similar final state was obtained in \cite{di_franco_nested_2008} 
using a combination of Ising-like interactions in two different directions in alternate sites. 
Our method is conceptually simpler, as it requires the same type of 
coupling throughout the chain.
On the other hand, Hamiltonians allowing \emph{almost}-perfect transfer
are easier to
implement experimentally, as they usually require a static tuning of a
single parameter 
\cite{banchi_ballistic_2013,banchi_optimal_2010,banchi_long_2011,bose_spin_2014}
rather then a full-engineering. 
We show that these models allow 
the generation a high entanglement 
between the ends of the chain, in principle even for $N{\to}\infty$ ({\em i.e.,} 
distance independent entanglement).

\section{Mapping between quench and state transfer} 
We consider a chain of $N$ spin-$\frac12$ particles coupled by the
following Hamiltonian 
\begin{align}
  \mathcal H_\Delta = 
  \frac{J}2\sum_{n=1}^N \left[j_n \left(\sigma_n^x\,\sigma_{n+1}^x 
  + \sigma_n^y\,\sigma_{n+1}^y\right) +\Delta \,
\sigma_n^z\,\sigma_{n+1}^z\right]~,
  \label{e.Hxx}
\end{align}
where $\sigma^\alpha_n$ are the Pauli matrices, $Jj_n$ are the coupling 
strengths ($J$ is the energy unit, while $j_n$ are adimensional) 
and $\Delta$ is the eventual anisotropy. We are interested in the
entanglement generation via the non-equilibrium evolution of the 
N\'eel initial state 
$\afm {=} \ket{\ua\da\ua\da{\dots}}$ under the XX Hamiltonian $\mathcal H_{\Delta=0}$. 
Formally this corresponds to a quench from $\Delta{=}\infty$ to $\Delta{=}0$, 
though the state $\afm$ can be prepared in different ways 
\cite{koetsier_achieving_2008}. 

We now prove a theorem which connects the many-body non-equilibrium evolution
following the quench to a state transfer problem. 
This connection will be then 
exploited to maximize the amount of generated entanglement. 
The Hamiltonian $\mathcal H_0$ can be mapped to a fermionic hopping
Hamiltonian via the Jordan-Wigner (JW) transformation
$c_n^\dagger{=}\prod_{m<n}({-}\sigma_m^z)\sigma_n^+$,
$\sigma_n^+{=}[\sigma_n^x{+}i\sigma_n^y]/2$: the new
operators obey fermionic anti-commutation relations
$\{c_n,c_m^\dagger\}{=}\delta_{nm}$, and $\mathcal H_0{=}\sum_{nm} A_{nm}
c_n^\dagger c_m$. The initial state $\afm$ has a fixed number of ``particles''
and $\mathcal H_0$ is quadratic and particle-conserving. 
Thus, owing to Wick's theorem, 
the evolved state is completely specified in the Heisenberg picture 
by the two-point correlation functions $ \mean{c_n^\dagger(t)c_m(t)}$ 
where $\mean{\cdot} {=} \bra{\rm AFM}\,{\cdot}\,\ket{\rm AFM}$. 
It is $\mean{c_n^\dagger(t)c_m(t)}{=} \sum_{ij} f_{ni}^*(t)f_{mj}(t) 
\mean{c_i^\dagger c_j}$ where $f{=}e^{-it A}$, in matrix notation.
By defining the sign matrix $S_{ij}{=}(-1)^{i+1}\delta_{ij}$ we find that $\mean{c_i^\dagger c_j}{=}(\delta_{ij}{+}S_{ij})/2$ and since $A$ is tridiagonal,
$SAS{=}{-}A$. Thus we obtain the following equality
\begin{align}
  \mean{c_n^\dagger(t)c_m(t)} = 
  \frac{\delta_{nm} + (-1)^{n+1} f_{nm}(2t)}2~,
  \label{e.id}
\end{align}
for any pair of sites $n,m$.
Before clarifying the implications of Eq.~\eqref{e.id}, 
we note that 
$f_{nm}(t)$ represents the probability amplitude for a fermionic quantum walker
to reach site $n$ at time $t$, starting from site $m$. 
In the single-particle sector, the fermionic nature of the walker does not
show up, and $|f_{nm}(t)|^2$ quantifies also the state transmission probability 
from $m$ to $n$ of spin $\ket\ua$ traveling in a
``sea'' of $\ket{\da}$ spins.
%
Therefore, if the Hamiltonian
\eqref{e.Hxx} for a particular set of couplings $\{j_n\}$ allows perfect 
single-excitation 
transfer from $m$ to $n$ at some time $t^*$, then starting from the many-body
$\afm$ initial state two fermions get completely delocalized among the
two distant sites $m,n$ at time $t^*/2$, i.e. half of the transmission time.
By taking into account the non-local relation between fermions and spins, in the
following we prove that at time $t^*/2$ the sites $m,n$ get maximally entangled
for particular Hamiltonians and pairs of spins.
Therefore, Eq.\eqref{e.id} relates the dynamical entanglement generation
from the quench to a simpler optimization of independent quantum walks.

We conclude this introductory discussion with a further comment on 
Eq.~\eqref{e.id}, to avoid confusions with Rabi-like dynamics.
Two qubits interacting via an XX Hamiltonian with coupling $J$, display perfect
state transfer ($\ket{\ua\da}\to\ket{\da\ua}$) after a time $t^*{=}\pi/(2J)$
and a Bell state generation 
($\sqrt 2\ket{\ua\da}\to\ket{\ua\da}{-}i\ket{\da\ua}$) 
after a time $t^*/2$.  However, the interpretation of Eq.~\eqref{e.id} as 
a long distance version of this behaviour is wrong. 
If a ballistic transfer happens on a time
$t^*$, the distant spins cannot be entangled by a single traveling particle 
after $t^*/2$: this would 
violate the Lieb-Robinson bound \cite{lieb_residual_1967}.
The physical explanation of the effect we describe is 
that each $\ket\ua$ in the initial states propagates in the two different
directions and contributes a certain amount of entanglement between 
the spins in its effective ``light cone'' \cite{calabrese_evolution_2005}.
However, a {\it single}, 
delocalized, hopping particle cannot produce maximal long distance entanglement.
The final amount of entanglement is due to the sum of different 
contributions given by each counter-propagating quasi-particle.  Therefore, 
the maximal entanglement generation presented in this article is a truly 
many-particle effect, which 
arguably depends on the particular alternation of spins $\ket\ua$ and $\ket\da$
in the N\'eel initial state  
\cite{wichterich_exploiting_2009,alkurtass_quench_2013},

\subsection{Fully engineered Hamiltonians for maximal block entropy}
\begin{figure}[t]
  \centering
  \includegraphics[width=.37\textwidth]{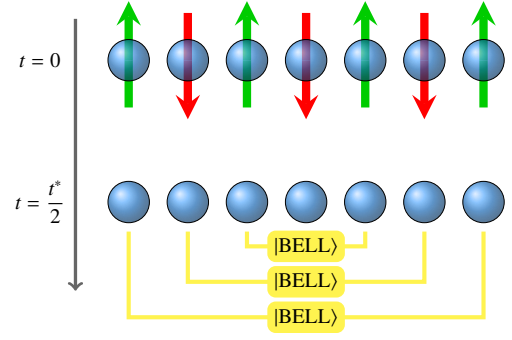}
  \caption{(Color online) Schematic picture for dynamical generation of 
    entanglement using a fully engineered XX Hamiltonian and the N\'eel 
  initial state.}
  \label{f.scheme}
\end{figure}

Perfect transmission can be achieved via 
engineered Hamiltonians such as
the XX chain with couplings $j_n {=} \sqrt{n(N-n)}/N$ \cite{christandl_perfect_2004,kay_review_2009}. 
In this engineered chain 
any walker starting from $n$ is exactly transmitted to its mirror symmetric
position $N{-}n{+}1$ at the transmission time $t^*{=}\pi N/(2J)$.
Thanks to the mapping Eq.~\eqref{e.id}, we show that, 
when the chain is initially set in the (separable) N\'eel state, 
after a time $t^*/2$ the state evolves into a maximal set 
of \emph{nested} Bell states. A schematic picture of 
this process is shown in Fig.~\ref{f.scheme}.
As $|f_{n,N{-}n{+}1}(t^*)|{=}1$ for any $n$,  it is  
$\ket{\Xi}{=}e^{-i \mathcal H_0 t^*/2}\afm \propto 
(c_1^\dagger {+}e^{i\alpha_1}c_N^\dagger) 
(c_2^\dagger {+}e^{i\alpha_2}c_{N{-}1}^\dagger)\cdots\ket{0}$, being
$\ket 0 $ the vacuum of the fermi operators and $\alpha_i$ some
defined phases.
In $\ket\Xi$ 
the spin in position $n$ is maximally entangled 
with the one in position $N{-}n{+}1$, as in Fig.~\ref{f.scheme}. 
Indeed, $c_n^\dagger{\propto}\sigma_n^+$ up to a phase which depends on the
number of spin $\ket\ua$ in position $m{<}n$. 
Proceeding 
recursively from the center of the chain, one can show that
$\ket{\Xi}{\propto} 
(\sigma_1^+ {+}e^{i\alpha'_1}\sigma_N^+) 
(\sigma_2^+ {+}e^{i\alpha'_2}\sigma_{N{-}1}^+)\cdots\ket{0}$,
for some new phases $\alpha'_n$.
In appendix A we prove that $\alpha'_n{=}\alpha$, so
the generated Bell states are 
$\ket{\rm BELL} {=} (\ket{\da\ua}{+}e^{i\alpha}\ket{\ua\da})/\sqrt2$, being
$\alpha{=}0$ for $N$ odd and $\alpha{=}\pi/2$ for $N$ even.
Accordingly, the left block of the chain is maximally
entangled with the right block.
\begin{figure}[t]
  \centering
  \includegraphics[width=.45\textwidth]{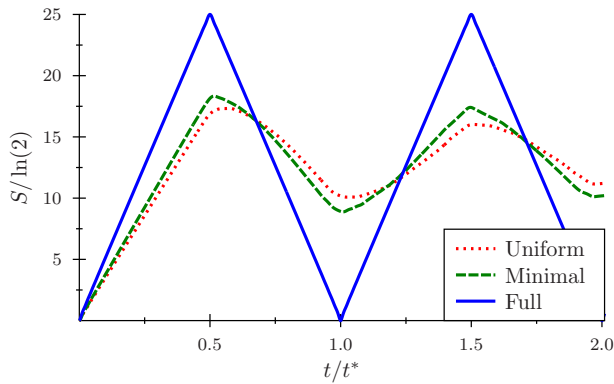}
  \caption{(Color online) Entropy dynamics for fully engineered, minimal engineered and
    uniform couplings. $N{=}51$.}
  \label{f.entro}
\end{figure}
The corresponding dynamics of the entropy is shown in Fig.~\ref{f.entro}.
Finally, our method is much more efficient to generate pairs of remote 
maximally entangled states. 
Indeed, the operational time of our protocol scales as $N$, while 
the generation of the same state via a composition of CNOT and SWAP gates would
require a number of operations of the order of $N{\times}N$.

\subsection{Minimally engineered model for distance-independent entanglement}
Minimally engineered chains 
\cite{banchi_optimal_2010,banchi_long_2011,banchi_ballistic_2013} 
 allows an optimal ballistic transmission between the chain boundaries.
Here only a single parameter is tuned, 
i.e. the couplings $j_1{=}j_{N{-}1}{=}j'$ at the ends. 
The other qubits are uniformly coupled with
$j_n{=}1/2$, $n{\neq}1,N{-}1$. When $j'$ is set to an optimal value 
${\propto} N^{ {-}1/6}$ the dynamics is ruled by the excitations with 
linear dispersion relation
\cite{banchi_long_2011}. Thus $|f_{N1}(t^*)|{\simeq} 1$ for a ballistic
transmission time $t^*{\approx} N/v$, being $v{\simeq}J$ the group 
velocity. 

Minimal engineering maximizes the transmission quality
between the chain ends, without optimizing the transmission between
other pairs. After the quench, we known from Eq.~\eqref{e.id}
that at time $t^*/2$ there is
an almost-maximally delocalized fermion $\tilde c$ between the two ends.
Since $c_N^\dagger{=}\sigma^+_N\,\Pi$, where
the parity $\Pi$ of the whole chain is a constant of motion and
$\Pi\afm{=}{\pm}\afm$, the fermion $\tilde c$ yields 
a highly entangled state between the ends of the chain. 
However, in this case $|f_{N1}(t^*)|{\neq}1$ because the fermion $\tilde c$ has
a non-zero probability of staying far from the ends, so the 
generated long-distance entangled state $\rho_{1,N}$ 
is not maximally entangled.  The amount of entanglement is quantified
by the fully entangled fraction 
$F(t){=}\max_{\ket e}\left\langle e| \rho_{1,N}(t) | e\right\rangle $,
where $|e\rangle$ is a maximally entangled states. 
When $F{>}\frac{1}{2}$ the state is purifiable 
(hence useful for teleportation \cite{bennett_mixed-state_1996}).
We found $F(t) {=}(1{+}|f_{N1}(2t)|)^2/4$.  
When the optimal $j'(N)$ is used, $|f_{N1}(t^*)|{\simeq} 85\%$ for
$N{\to}\infty$ \cite{banchi_long_2011}.
Therefore the generated entangled state is almost distance-independent as 
$F{>} 85\%$ even in the infinite site limit. 
Minimal engineering 
increases the resulting entanglement significantly compared to
the uniform case \cite{wichterich_exploiting_2009}. 
For smaller chains $F$ is fairly larger 
(see appendix B) and, e.g.,
for $N{=}25$ it is $F{=}97\%$.

\subsection{Quench from other initial states} 
All the results
discussed so far can be obtained also with the following Hamiltonian
\begin{align}
  \mathcal H'_B = 
  \frac{J}2\sum_{n=1}^N j_n \left(\sigma_n^x\,\sigma_{n+1}^x 
  - \sigma_n^y\,\sigma_{n+1}^y\right) -B \sum_n \sigma_n^z~,
  \label{e.Hdq}
\end{align}
by quenching the magnetic field from $B{=}\infty$ to $B{=}0$. 
The initial state in this case is $\fm{=}\ket{\ua\ua\ua{\dots}}$ and
$e^{-i\mathcal H'_0t}\fm {=} \prod_{n\text{ even}}\sigma_n^x\,
e^{-i\mathcal H_0t}\afm$. 
As $\prod_{n\text{ even}}\sigma_n^x$ is a product of local rotations, the 
states 
$e^{-i\mathcal H_0t}\afm$ and $e^{-i\mathcal H'_0t}\fm$ share the same amount
of entanglement. 

In appendix C we study also the quench from a different initial state
(series of nearest-neighbor Bell states) 
that might be easier to generate in some experimental setups
\cite{trotzky_controlling_2010}.

\section{Experimental proposals}
\subsection{NMR-based implementation}
Pulsed control techniques in Nuclear Magnetic Resonance (NMR) 
have reached a high degree of maturity \cite{negrevergne_benchmarking_2006} 
and provide a platform to observe 
quantum dynamics and state transfer in 
spin chains \cite{cappellaro_implementation_2014,rao_simulation_2013}. 
The natural dipolar interactions between the nuclear spins 
can be tuned \cite{zhang_nmr_2009,ajoy_quantum_2013}
and an effective ``double-quantum'' Hamiltonian \eqref{e.Hdq} obtained. 
In particular, a suitable pulse sequence to engineer 
the coupling strengths according to $j_n{=}\sqrt{n(N-n)}/N$
has been recently proposed \cite{ajoy_quantum_2013}.
At $t^*{=}\pi N/(2J)$, where $J{\approx}5\,{\rm KHz}$,
a nearly lossless state transfer is expected for chains as 
long as $N{=}25$ \cite{ajoy_quantum_2013}. 
Owing to our mapping, a near perfect generation of 
{\it nested} Bell states is expected
when an initially polarized state $\fm$ evolves under $\mathcal H_0'$ 
for a time $t^*/2$. 
In the high temperature regime, a \emph{pseudo-pure} initial state 
$\rho_{\rm pp}{=}\frac\zeta{2^N}\openone{+}(1{-}\zeta)\ket{\rm FM}
\bra{\rm FM}$ may be implemented 
using standard averaging techniques if enough control on the spins is
available
\cite{cory_ensemble_1997,gershenfeld_bulk_1997,knill_theory_1997}.
The initialization error $\zeta$ leads to 
$F{=}1{-}3\zeta/4$. When $\zeta$ is low we have actual etanglement, 
while when it is high the correlations enable one to 
verify the protocol. 
The spins at the ends can be read out 
\cite{kaur_initialization_2012} exploiting their peculiarity of having
just one nearest neighbor.
The main error sources are typically  pulse errors and 
intra-chain interactions
\cite{zhang_nmr_2009}. For simplicity, 
we model these errors as an imperfect filtered engineering
\cite{ajoy_quantum_2013}
of $\mathcal H'_B$. 
We consider $\tilde{\mathcal H}'{=}J\sum_{n{\neq}m}A_{nm} 
(\sigma_n^x\sigma_m^x{-} \sigma_n^y\sigma_m^y)$, 
$A_{nm}{=}j_n(\delta_{n,m{+}1}{+}\delta_{n{+}1,m})
{+}\epsilon\, b_{nm} F_{nm}$ where $\epsilon$ is the error strength, 
$b_{nm}$ models the long range interactions and $F$ models the
imperfect filtering. To include the effect of a nearby chain, we consider
two parallel chains coupled via dipolar interaction $b_{nm}$,
where the interchain distance is three times the distance between
intrachain spins \cite{zhang_nmr_2009}. 
\begin{figure}[t]
  \centering
  \includegraphics[width=.35\textwidth]{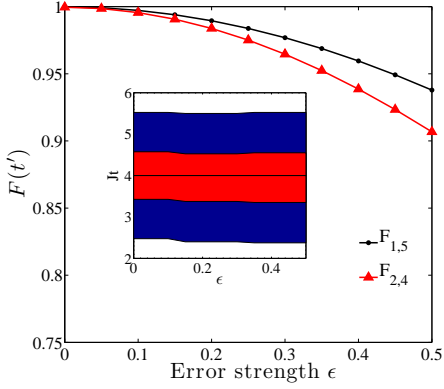}
  \caption{ (Color online) 
    Fully entangled fraction $F_{n,N{-}n{+}1}(t')$ for pairs of mirror
  symmetric spins at the time $t'{\simeq}t^*/2$ which
  maximizes $F_{1,N}$, $N{=}5$. For $N{=}5$ fully and minimally engineered
  chains coincide.
  The pulse error $\epsilon$ is defined in the text. 
  An average over $100$ realizations of the imperfections is considered. 
  Inset: $t'$ (black line), time where $F_{1,5}{\ge}0.9F_{1,5}(t')$ 
  (red or light gray area), time where $F_{1,5}{\ge}0.5F_{1,5}(t')$ 
  (blue or dark gray area).
}
  \label{f.NMR_error}
\end{figure}
The elements $F_{nm}{\in}[{-}1,1]$ are chosen at random.  
The results are shown in Fig.~\ref{f.NMR_error}.

\subsection{Ion Traps Implementation}
\begin{figure}[t]
  \centering
  \includegraphics[width=.45\textwidth]{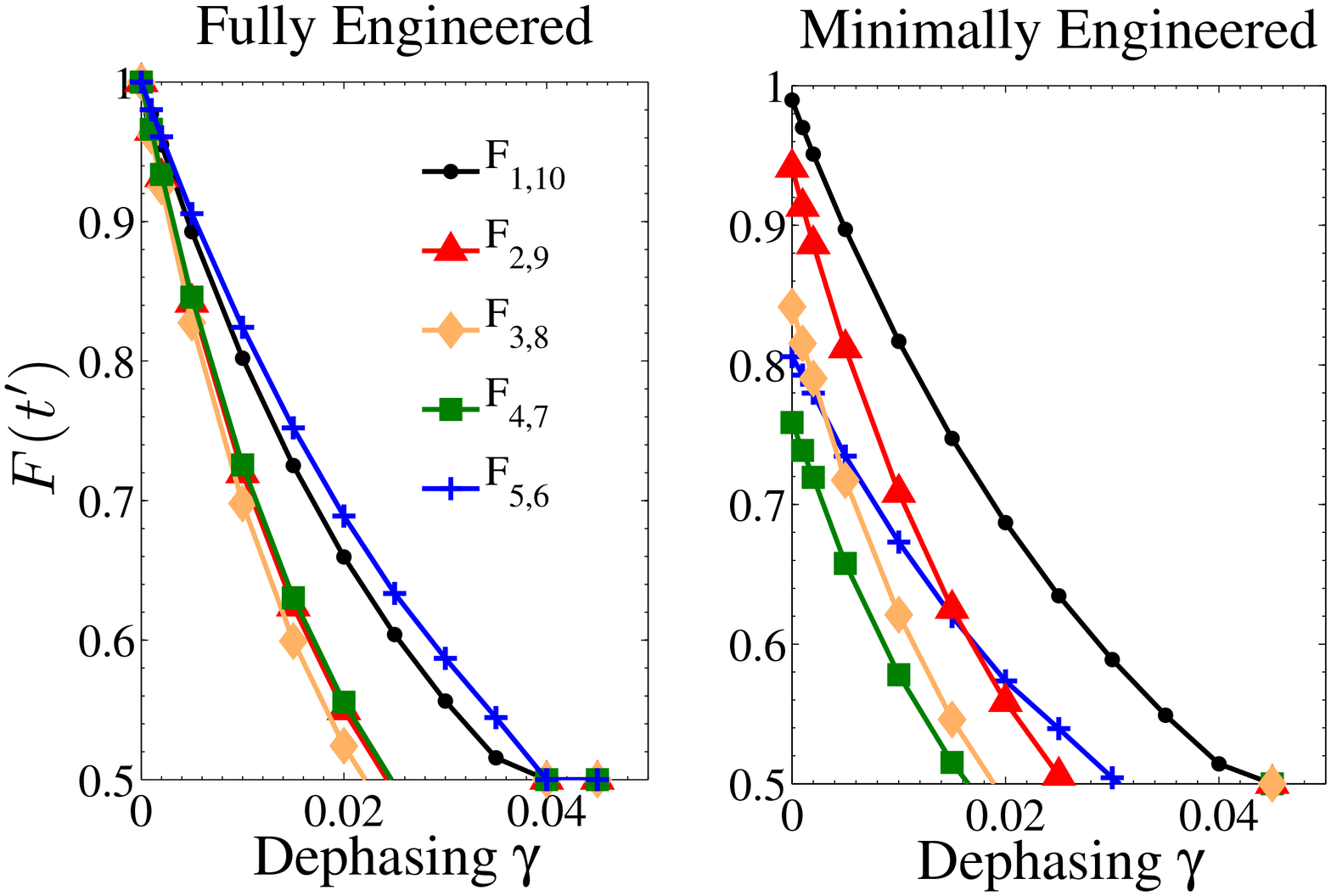}
  \caption{(Color online) 
    $F_{n,N{-}n{+}1}(t')$ as in Fig.~\ref{f.NMR_error}.  
  Fully engineered (left) and minimally engineered 
  (right) chains for $N{=}10$. 
  Decoherence is modeled by the master equation 
  $\dot\rho{=}{-}i[\mathcal H_0,\rho]{+}J\gamma\sum_i(\sigma_i^z\rho\sigma_i^z
  {-}\rho)$ with dephasing rate $\gamma$ \cite{zippilli_adiabatic_2013}.  }
  \label{f.deco}
\end{figure}

Ion traps represent a promising implementation of a quantum computer where two
internal hyperfine states of each trapped ion implement a qubit
\cite{mc_hugh_quantum_2005,johanning_quantum_2009,khromova_designer_2012}.
Ising-like coupling between ions can be induced using a magnetic field gradient
\cite{wunderlich_conditional_2002,wunderlich_two-dimensional_2009}. 
In addition, the magnetic field gradient
causes local frequency shifts allowing for addressing and
manipulation of individual ions using microwave pulses. Segmented
microstructured traps provide the possibility of tailoring the couplings 
via local trapping potential, 
thus allowing for suppression of long-range couplings
\cite{wunderlich_two-dimensional_2009}. 
Alternatively, spin-spin coupling can be generated by laser-induced forces
\cite{porras_effective_2004,richerme_non-local_2014,jurcevic_observation_2014}.
An effective XX Hamiltonian
can be implemented either with a large transverse 
magnetic field \cite{jurcevic_observation_2014} or via 
fast sequential applications of Ising evolution in two orthogonal directions 
\cite{zippilli_adiabatic_2013}. 
The main sources of error here are long-range
interactions and dephasing. 
The robustness of our scheme against decoherence is studied in 
Fig.~\ref{f.deco} in terms of the dephasing rate $J\gamma$. 
As $J{\approx} 1 \,{\rm kHz}$ \cite{zippilli_adiabatic_2013} we find that
with a decoherence time $1/(J\gamma){\simeq}100{\rm ms}$ the entangled
fraction $F$ can be higher than $80\%$ for $N=10$ (dynamical decoupling as in 
Ref.~\cite{piltz_protecting_2013} can be used as our scheme is invariant 
to $\pi$ pulses). 
To study the effect of long-range interactions 
we use the approximation
$j_{nm} {=} (\omega_n^2\omega_m^2 |n-m|^3)^{-1}$, 
valid when the Coulomb interaction is a perturbation of the trapping potential
(see e.g. \cite{porras_effective_2004,kim_entanglement_2009});
$\omega_n$ is the frequency of the local trap, chosen
such that $j_{n,n+1}{=}j_n$. The result 
is displayed in Table.~\ref{LongRangeTable}.

\begin{table}[t]
\begin{tabular}{ | c | c | c | c | c | c | }
  \hline
  $F_{n,N-n+1}(t')$ & $n=1$ & $n=2$ & $n=3$ & $n=4$ & $n=5$ \\
\hline  
 Fully engineered 
 & 0.88 & 0.82 & 0.83 & 0.85 & 0.88 \\
\hline  
 Minimally engineered 
 & 0.88 & 0.77 & 0.69 & 0.57 & 0.60 \\
  \hline
\end{tabular}
\footnotesize
  \caption{Fully entangled fraction with long range interactions, where
    $t'{\simeq} t^*/2$, $N{=}10$.}
\label{LongRangeTable}
\end{table}


\subsection{Optical lattice implementation}
In optical 
lattices, the qubit is encoded into two hyperfine states of a
neutral atom 
\cite{trotzky_controlling_2010,lubasch_adiabatic_2011,trotzky_time-resolved_2008}. 
Nearly all requirements for the implementation of our scheme have been met is Refs. \cite{fukuhara_microscopic_2013,fukuhara_quantum_2013} -- for example, lines of 10-14 atoms without defects
forming a spin chain can be post-selected using a quantum gas microscope, which
can also be used to readout individual atomic state to verify the generated
entanglement. The N\'eel state initialization should be possible by preparing a
spin polarized state and flipping alternate spins by combined action of light
from a spatial light modulator (SLM) and a resonant microwave pulse in exactly
the same manner as a single \cite{fukuhara_quantum_2013} and double
\cite{fukuhara_microscopic_2013} rows of spins were selectively flipped in
recent experiments. 
After the initialization, one should suddenly (in ${\ll} 0.1$s) change the 
laser field from the SLM to that with an appropriate intensity profile
\cite{clark_efficient_2005} to generate the effective coupling $j_n$,
as well as to provide a hard wall end to the lattice \cite{banchi_nonperturbative_2011}. 
An effective XX Hamiltonian with $J{\approx}50{\rm Hz}$, 
should be possible by tuning (e.g. by Feshbach resonance) 
interspecies and intraspecies interactions
\cite{duan_controlling_2003}. 
On the other hand, 
experimental imperfectnesses can introduce
spurious interactions as an effective 
anisotropic $J_z\sigma^z_{n}\sigma^z_{n+1}$ term. 
We found that the entangling scheme is robust against these imperfections, 
with $90\%$ of the 
entanglement preserved for upto $J_z{\approx} 0.35 j_n$.

\section{Conclusions}
We have mapped the correlations 
between two spins $m,n$ 
produced by specific quenches at $t$ to a different dynamics, namely
a spin transfer between sites $m$ and $n$ at $2t$. 
This nontrivial mapping allows one to generate
both unbounded block-block and useful spin-spin entanglement from a global quench in contrast to much condensed mater literature. We found that a full or a minimal engineering of
the couplings simultaneously generates many Bell
states (maximal block entropy) or distance-independent
entanglement. 
Our method is much more efficient than a composition of CNOT and SWAP gates to generate a full set of Bell pairs: the former scales as $t{\sim}N$ while the latter requires $t{\sim}\mathcal O(N^2)$. 
Finally, we have studied the feasibility of realizations
in NMR, ion traps and 
ultracold atoms in optical lattices.
Given the sub-decoherence operational times $({\approx}{\rm ms})$
and the low control required, our method should be competitive
for linking quantum registers.

\section{Acknowledgements}%
Discussions with Abolfazl Bayat, Paola Cappellaro,
Takeshi Fukuhara and Christof Wunderlich 
are warmly acknowledged. 
SB and LB are supported by the ERC grant PACOMANEDIA. 
BA is supported by King Saud University.

\appendix

\section{Proof of maximal generation of Bell states }
In this section we show that, using a full engineered chain one can
generate a maximal set of long-distance bell states as in Fig.~\ref{f.scheme}.
For small enough $N$, one can evaluate efficiently the evolved state 
exploiting the Fermionic nature of the XX evolution. The time evolved 
state can indeed be written in a determinant form \cite{banchi_ballistic_2013}
\begin{align}
  e^{-i \mathcal H_0 t} \ket{m_1, m_2, \dots} = 
  \sum_{\ell_1 < \ell_2 < \dots} \det\left[ f_{(\ell_1,\ell_2,\dots),
  (m_1,m_2,\dots)}(t)\right]\ket{\ell_1,\ell_2,\dots}
  \label{e.det}
\end{align}
where $\ket{m_1, m_2, m_3, \dots}$ represents a 
state with a spin $\ket\da$ in positions $m_1<m_2<m_3 \dots$, and 
$f_{(\ell_1,\ell_2,\dots),
    (m_1,m_2,\dots)}(t)$ is the submatrix of $f$ with rows specified by
    the index vector $(\ell_1,\ell_2, \dots),$ 
    and columns specified by $(m_1,m_2,\dots)$. 

The matrix elements $f_{nm}(t)$ can be calculated 
analytically. One can show \cite{christandl_perfect_2004} indeed that
\begin{align}
  f_{n'n}(t) = \bra{r'} e^{-i\frac{2 t}{N} S^{(s)}_x} \ket r~,
  \label{e.Sx}
\end{align}
where
\begin{align}
  s&=\frac{N-1}2&r&=-s+n-1 &r'&=-s+n'-1~.
  \label{e.Sxdef}
\end{align}
Namely, the engineered quantum walk is equivalent to 
the rotation of an effective spin-$\frac{N-1}2$ along the $x$ direction. 
Owing to this equivalence the time evolution can be expressed in terms of
the Wigner $\mathcal D$ matrix \cite{biedenharn_angular_1981}: 
\begin{align}
    f_{n'n}(t) = \mathcal D^{(s)}_{m'm}\left(\frac\pi2, 
      -\frac{2 t}{N}, -\frac\pi2\right) ~,
      \label{e.UD}
\end{align}
where the definitions \eqref{e.Sxdef} have been used. The matrix elements
of the Wigner matrix are well known. For instance, from
\begin{align}
  f_{n'n}(t^*) = (-i)^{N-1} \; \delta_{n',N-n+1}~,
  \label{e.ftstar}
\end{align}
it is now clear that the engineered chain acts as a perfect mirror 
after a time $t^*=N \pi/2$. However, the analytical expression for $t=t^*/2$ is 
not so simple.

We prove the structure of Fig.~\ref{f.scheme} thanks to
Eq.~\eqref{e.id}. Indeed, owing to Wick's theorem, the evolved state 
can be completely specified by its two point correlation function, up to
a global phase.
Let us set $N=2M-1$ for odd $N$ or $N=2M$ for even $N$ and 
\begin{align}
  e^{-it \mathcal H_0}\afm= \prod_{k=1}^M\left(\sum_n U_{nk}(t) 
  c_n^\dagger\right)
  \ket 0~.
  \label{e.fermiU}
\end{align}
Clearly one can set $U_{n1} = f_{n1}^*$, $U_{n2} = f_{n3}^*$, 
$U_{n3} = f_{n5}^*$, etc., so that the result is \eqref{e.det}. However, 
there is some arbitrary freedom in choosing $U$: also with a different
$M\times N$ matrix $U$ the result can still be that of \eqref{e.det}. 
We exploit this arbitrariness in order to simplify the derivation. 
By calculating $R_{nm}(t)= \mean{c_n(t)c_m^\dagger(t)}$ with the ansatz
\eqref{e.fermiU} one finds
\begin{align}
  R_{nm}(t) = \det\left[U^{(n)}{}^\dagger \,U^{(m)}\right]~,
  \label{e.defR}
\end{align}
where $U^{(m)}$ is built from $U$ by adding the column vector 
$e^{(m)}$ which has only one non-zero element, 
$(e^{(m)})_m = 1$. Then $R_{nm}$ can be written as a determinant of a 
$(M+1)\times (M+1)$ matrix in a block form
\begin{align}
  R_{nm} = \det\begin{pmatrix}
    \delta_{nm} & U_{m\cdot} \\
    U_{n\cdot}^\dagger & U^\dagger\,U
  \end{pmatrix}
\end{align}
where $U_{n\cdot}$ is the $n$-th row of $U$. Using the well known identity
$\det\begin{pmatrix}
  A & B\\C&D
\end{pmatrix} = \det D\det(A-BD^{-1}C)$ and exploiting the fact
that $U^\dagger\,U$ is a submatrix of a unitary matrix one obtains
\begin{align}
  R_{nm} = \delta_{nm} - \left(U\,U^\dagger\right)_{nm}~,
  \label{e.RU}
\end{align}

Thanks to Eq.\eqref{e.id}, one can write $R_{nm}(t) = [\delta_{nm}+(-1)^m
f_{nm}(2t)]/2$. Using \eqref{e.ftstar} 
\begin{align}
  R_{nm}(t^*/2) = \frac{\delta_{nm} + (-1)^m (-i)^{N-1} \; \delta_{m,N-n+1}}2~.
  \label{e.Rfinal}
\end{align}
By imposing that \eqref{e.Rfinal} and \eqref{e.RU} are equal one finds the
simple solution 
\begin{align}
  U_{nk}(t^*/2) = \begin{cases}
    \frac{\delta_{n,k} + e^{i\alpha_{n}}\delta_{n,N-k+1}}{\sqrt 2} & 
    \text{~if~} k\neq N-k+1,\\
    1 &\text{~if~} k= N-k+1,
  \end{cases}
  \label{e.Ufinal}
\end{align}
where $e^{i \alpha_{n}} = i^{N-1}(-1)^{n-1} $. Therefore, one obtains that
\begin{align}
\ket\Xi=e^{-i \mathcal H_0 t^*/2}\afm =
\left(\frac{c_1^\dagger {+}e^{i\alpha_1}c_N^\dagger}{\sqrt 2}\right) 
\left(\frac{c_2^\dagger {+}e^{i\alpha_2}c_{N{-}1}^\dagger}{\sqrt 2}\right)\cdots\ket{0}
  \label{e.XifermiEven}
\end{align}
for $N$ even and 
\begin{align}
\ket\Xi=
\left(\frac{c_1^\dagger {+}e^{i\alpha_1}c_N^\dagger}{\sqrt 2}\right) 
\left(\frac{c_2^\dagger {+}e^{i\alpha_2}c_{N{-}1}^\dagger}{\sqrt
2}\right)\cdots c^\dagger_{(N+1)/2}\ket{0}
  \label{e.XifermiEven}
\end{align}
for $N$ odd. Going back into the spin representation, exploiting 
the anti-commutation relations, one obtains that $\ket\Xi$ consists of
a product of maximally entangled states, as in Fig.~\eqref{f.scheme}
\begin{align}
  \ket\Xi_{\rm even} &=\prod_{k=1}^{\frac N 2} 
  \left(\frac{\ket{\ua\da}_{k,\tilde k}-
  i\ket{\da\ua}_{k,\tilde k}}{\sqrt{2}}\right)
  \\
  \ket\Xi_{\rm odd} &=\ket{\ua}_{\frac{N+1}2}\;\prod_{k=1}^{\frac{N-1}2}
  \left(\frac{\ket{\ua\da}_{k,\tilde k}+
  \ket{\da\ua}_{k,\tilde k}}{\sqrt{2}}\right)
\end{align}
where $\tilde k=N-k+1$. A schematic picture of the resulting state is
drawn in Fig.~\ref{f.scheme} in the main text.

\section{Distance-independent entanglement generation}
\begin{figure}[t]
  \centering
  \includegraphics[width=.45\textwidth]{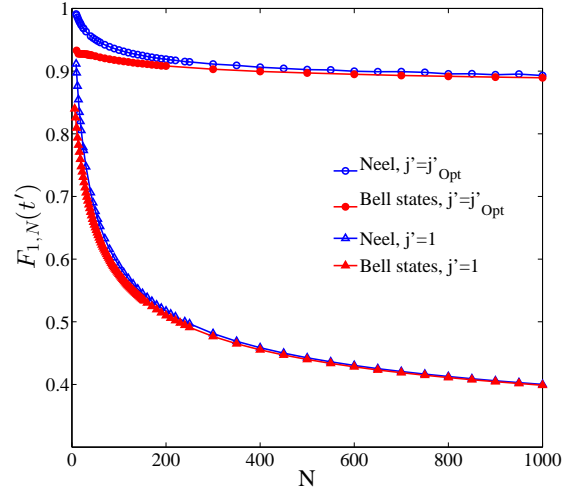}
  \caption{(Color online) Maximum $F_{1,N}$ for the minimally engineered Hamiltonian versus the chain length.}
  \label{F_vs_N}
\end{figure}

\begin{figure}[t]
  \centering
  \includegraphics[width=.45\textwidth]{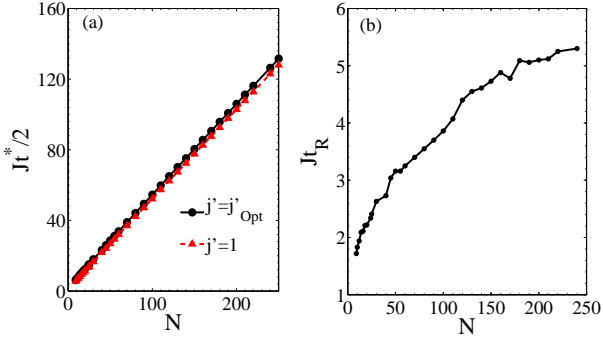}
  \caption{(Color online) (a) Time at which the first peak of $F_{1,N}$ occurs versus the chain length for the minimally engineered Hamiltonian and N\'eel initial state (b) Reading time defined as the width of $F_{1,N}(t)$ at half the maximum.}
  \label{time}
\end{figure}

As stated in the main text, the Hamiltonian $\mathcal H_0$ can be
expressed as a JW-fermionic hopping model 
\begin{equation}
\mathcal H_0=\sum_{n,m}\hat{c}_{n}^{\dagger}  A_{n,m} \hat{c}_{m} \,,
\label{HA}
\end{equation}
where $A_{n,m}=j_n(\delta_{n,m+1}+\delta_{n+1,m})$. The hopping matrix $A$ can be diagonalized with an orthogonal matrix $g$ where $\sum_{i,j=1}^{N} g_{k,i} A_{i,j} g_{l,j} = E_k \delta_{k,l}$. The diagonalization was done analytically in Ref.\cite{banchi_spectral_2013}. Hence the fermionic operators in the Heisenberg picture are found to be $\hat{c}_{k}(t)=\sum_{l} f_{k,l} \hat{c}_{l}(0), \quad \hat{c}_{k}^{\dagger}(t)=\sum_{l} f_{k,l}^{*} \hat{c}_{l}^{\dagger}(0)$ where $f_{k,l}(t)= \sum_{m=1}^{N} g_{m,k} g_{m,l} e^{-i E_m t}$.\\
In the basis $\left\{\left|\uparrow\uparrow\right\rangle,\left|\uparrow\downarrow\right\rangle,\left|\downarrow\uparrow\right\rangle,\left|\downarrow\downarrow\right\rangle\right\}$, the only non-vanishing elements of the density matrix of the distant ends spins $\rho_{1,N}$ are
\begin{equation}
\rho_{1,N} =\left( \begin{array}{cccc}
\rho_{11} \\
& \rho_{22} & \rho_{23} \\
   &  \rho_{32} & \rho_{33} \\
   &   &  & \rho_{44} \\
\end{array}\right) \, .
\end{equation}
Defining
\begin{equation}
\alpha=\left\langle\hat{c}_{1}^{\dagger}(t)\hat{c}_{1}(t)\right\rangle\, ,
\beta=\left\langle\hat{c}_{N}^{\dagger}(t)\hat{c}_{N}(t)\right\rangle \, ,
\gamma=\left\langle\hat{c}_{1}^{\dagger}(t)\hat{c}_{N}(t)\right\rangle \, ,
\end{equation}
we find the density matrix elements to be
\begin{align}
\nonumber\rho_{11} =&\alpha \beta -|\gamma|^2, &\rho_{44}&= (1-\alpha)(1-\beta) -|\gamma|^2 \, , \\
\rho_{22}=&\alpha (1-\beta) + |\gamma|^2, &\rho_{33}&=\beta (1-\alpha) + |\gamma|^2 \, ,\\
\nonumber\rho_{23}=& (-1)^{M+1}\gamma^{*}, &\rho_{32}&= (-1)^{M+1}\gamma \, ,
\end{align}
where the two point correlation function is given by
\begin{equation}
\left\langle \hat{c}_{i}^{\dagger}(t) \hat{c}_{j}(t)\right\rangle=\sum_{m=odd} f_{i,m}^{*}(t) f_{j,m}(t) \left\langle \hat{c}_m^{\dagger}(0) \hat{c}_m (0) \right\rangle \, ,
\end{equation}
and $M$ is the number of up spins in the initial state, i.e. $N/2$ for even $N$ and $(N+1)/2$ for odd $N$.
Having the analytic expression for the density matrix, we now evaluate the entanglement between spins 1 and $N$ using the fully entanglement fraction $F$.
For $\rho_{1,N}$ we find that
\begin{eqnarray}
\nonumber F_{1,N}=\max\Bigg\{\frac{\alpha\beta+(\alpha-1)(\beta-1)}{2}-|\gamma|^2,\\
\frac{\alpha+\beta}{2}-\alpha\beta+|\gamma|(1+|\gamma|)\Bigg\} \, .
\label{maxf}
\end{eqnarray}
Substituting for $\alpha, \beta,$ and $\gamma$ at half the transmission time into eq.(\ref{maxf}) we get the fully entangled fraction
\begin{equation}
F_{1,N}(t^*/2)=\frac{1}{4} \left(1+|f_{1,N}(t^*)|^2 \right) \, .
\end{equation}
From Ref.~\cite{banchi_long_2011}, $|f_{1,N}(t^*)|$ was found to approach an asymptotic value of $0.8469$ for large $N$ and hence the asymptotic value of the generated entanglement $F$ in this work is $0.8528$. The optimal time for entanglement generation from a quench would be half the time required for state transfer and hence $t^*/2=\frac{1}{2J}(N+2.29N^{1/3})$. The resulting $F_{1,N}$ is shown in Fig.~\ref{F_vs_N} and compared with the case of a homogeneous XX Hamiltonian \cite{wichterich_exploiting_2009}. The time $t^*/2$ is shown in Fig.~\ref{time}(a). Fig.~\ref{time}(b) shows the reading time defined as the time interval where $F_{1,N}>F(t')/2$.

\begin{figure}[t]
  \centering
  \includegraphics[width=.45\textwidth]{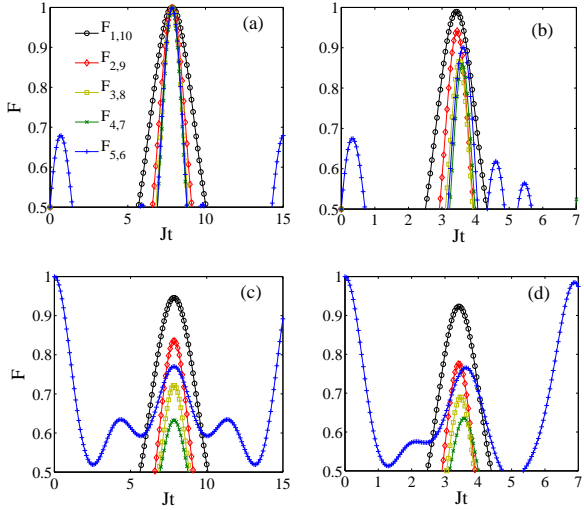}
  \caption{(Color online) Fully entangled fraction for $N=10$ for (a,b) N\'eel initial state with the fully engineered and minimally engineered Hamiltonian, respectively. (c,d) Series of Bell states initial state with the fully engineered and minimally engineered Hamiltonian, respectively.}
  \label{FractionN10}
\end{figure}

Fig.~\ref{FractionN10}(a,b) shows the fully entangled fraction evaluated numerically for the mirror-symmetric spins in a chain of length $N=10$ for the fully engineered and minimally engineered Hamiltonian considered in this article.

\section{Series of Bell States}
We also consider the following product of Bell states as an initial state,
\begin{equation}
\left|\psi\right\rangle=\mathop{\otimes}_{k=1}^{N/2} \left|\psi_{k,k+1}\right\rangle = \mathop{\otimes}_{k=1}^{N/2} \left(\left|\uparrow_{k} \downarrow_{k+1}\right\rangle - \left|\downarrow_{k} \uparrow_{k+1}\right\rangle \right).
\end{equation}
In this case, the density matrix elements $\rho_{1,N}$ is found to be
\begin{eqnarray}
\nonumber\rho_{11} &=&\alpha \beta -|\gamma|^2, \,\,\,\,\,\,\qquad\rho_{44}= (1-\alpha)(1-\beta) -|\gamma|^2 \\
\rho_{22}&=&\alpha (1-\beta) + |\gamma|^2, \,\,\,\rho_{33}=\beta (1-\alpha) + |\gamma|^2\\
\nonumber\rho_{23}&=& \delta^{*}, \qquad\qquad\qquad\,\rho_{32}=\delta \, ,
\end{eqnarray}
where we defined
\begin{eqnarray}
\nonumber\alpha&=&\left\langle\hat{c}_{1}^{\dagger}(t)\hat{c}_{1}(t)\right\rangle\! ,
\beta=\left\langle\hat{c}_{N}^{\dagger}(t)\hat{c}_{N}(t)\right\rangle \! ,
\gamma=\left\langle\hat{c}_{1}^{\dagger}(t)\hat{c}_{N}(t)\right\rangle \! , \\
\delta&=&\sum_{l,m} f_{1,l}^{*}(t) f_{N,m}(t) \left\langle \hat{c}_m \hat{c}^{\dagger}_l \otimes_{i=1}^{N}\left(-\sigma^z_i\right) \right\rangle \, .
\end{eqnarray}
The two point correlation functions are given by $\left\langle \hat{c}_{i}^{\dagger}(t) \hat{c}_{j}(t)\right\rangle=\sum_{l,m} f_{i,l}^{*}(t) f_{j,m}(t) \left\langle \hat{c}^{\dagger}_l \hat{c}_m \right\rangle$ where
\begin{equation}
\left\langle \hat{c}^{\dagger}_l \hat{c}_m \right\rangle=\left\{
	\begin{array}{ll}
		\frac{1}{2},  & l=m\\
		-\frac{1}{2}, & \left|l-m\right|=1\\
		0, & otherwise
	\end{array}
\right.
\end{equation}
and
\begin{equation}
\left\langle \hat{c}_l \hat{c}^{\dagger}_m \otimes_{i=1}^{N}\left(-\sigma^z_i\right) \right\rangle=\left\{
	\begin{array}{ll}
		\frac{1}{2}(-1)^{N/2},  & l=m\\
		\frac{1}{2}(-1)^{N/2}, & l=odd, m=l+1\\
		\frac{1}{2}(-1)^{N/2}, & l=even, m=l-1\\
		0, & otherwise
	\end{array}
\right.
\end{equation}
we then find the fully entangled fraction of $\rho_{1,N}$

\begin{eqnarray}
\nonumber F_{1,N}(t)=\max\Bigg\{\frac{\alpha\beta+(\alpha-1)(\beta-1)}{2}-|\gamma|^2,\\ \frac{\alpha+\beta}{2}-\alpha\beta+|\gamma|^2\pm \delta\Bigg\} \, .
\label{maxfSinglets}
\end{eqnarray}

The resulting $F_{1,N}$ are shown in Fig.~\ref{F_vs_N} for two initial states: (i) N\'eel state and (ii) series of Bell states.
\noindent Fig.~\ref{FractionN10} shows the fully entangled fraction evaluated numerically for the mirror-symmetric spins in a chain of length $N=10$ for the fully engineered and minimally engineered Hamiltonian and the two initial states considered in this work.


\begin{thebibliography}{76}%
\makeatletter
\providecommand \@ifxundefined [1]{%
 \@ifx{#1\undefined}
}%
\providecommand \@ifnum [1]{%
 \ifnum #1\expandafter \@firstoftwo
 \else \expandafter \@secondoftwo
 \fi
}%
\providecommand \@ifx [1]{%
 \ifx #1\expandafter \@firstoftwo
 \else \expandafter \@secondoftwo
 \fi
}%
\providecommand \natexlab [1]{#1}%
\providecommand \enquote  [1]{``#1''}%
\providecommand \bibnamefont  [1]{#1}%
\providecommand \bibfnamefont [1]{#1}%
\providecommand \citenamefont [1]{#1}%
\providecommand \href@noop [0]{\@secondoftwo}%
\providecommand \href [0]{\begingroup \@sanitize@url \@href}%
\providecommand \@href[1]{\@@startlink{#1}\@@href}%
\providecommand \@@href[1]{\endgroup#1\@@endlink}%
\providecommand \@sanitize@url [0]{\catcode `\\12\catcode `\$12\catcode
  `\&12\catcode `\#12\catcode `\^12\catcode `\_12\catcode `\%12\relax}%
\providecommand \@@startlink[1]{}%
\providecommand \@@endlink[0]{}%
\providecommand \url  [0]{\begingroup\@sanitize@url \@url }%
\providecommand \@url [1]{\endgroup\@href {#1}{\urlprefix }}%
\providecommand \urlprefix  [0]{URL }%
\providecommand \Eprint [0]{\href }%
\providecommand \doibase [0]{http://dx.doi.org/}%
\providecommand \selectlanguage [0]{\@gobble}%
\providecommand \bibinfo  [0]{\@secondoftwo}%
\providecommand \bibfield  [0]{\@secondoftwo}%
\providecommand \translation [1]{[#1]}%
\providecommand \BibitemOpen [0]{}%
\providecommand \bibitemStop [0]{}%
\providecommand \bibitemNoStop [0]{.\EOS\space}%
\providecommand \EOS [0]{\spacefactor3000\relax}%
\providecommand \BibitemShut  [1]{\csname bibitem#1\endcsname}%
\let\auto@bib@innerbib\@empty
\bibitem [{\citenamefont {Bennett}\ \emph {et~al.}(1993)\citenamefont
  {Bennett}, \citenamefont {Brassard}, \citenamefont {Crépeau}, \citenamefont
  {Jozsa}, \citenamefont {Peres},\ and\ \citenamefont
  {Wootters}}]{bennett_teleporting_1993}%
  \BibitemOpen
  \bibfield  {author} {\bibinfo {author} {\bibfnamefont {C.~H.}\ \bibnamefont
  {Bennett}}, \bibinfo {author} {\bibfnamefont {G.}~\bibnamefont {Brassard}},
  \bibinfo {author} {\bibfnamefont {C.}~\bibnamefont {Crépeau}}, \bibinfo
  {author} {\bibfnamefont {R.}~\bibnamefont {Jozsa}}, \bibinfo {author}
  {\bibfnamefont {A.}~\bibnamefont {Peres}}, \ and\ \bibinfo {author}
  {\bibfnamefont {W.~K.}\ \bibnamefont {Wootters}},\ }\href@noop {} {\bibfield
  {journal} {\bibinfo  {journal} {Physical Review Letters}\ }\textbf {\bibinfo
  {volume} {70}},\ \bibinfo {pages} {1895} (\bibinfo {year}
  {1993})}\BibitemShut {NoStop}%
\bibitem [{\citenamefont {Moehring}\ \emph {et~al.}(2007)\citenamefont
  {Moehring}, \citenamefont {Maunz}, \citenamefont {Olmschenk}, \citenamefont
  {Younge}, \citenamefont {Matsukevich}, \citenamefont {Duan},\ and\
  \citenamefont {Monroe}}]{moehring_entanglement_2007}%
  \BibitemOpen
  \bibfield  {author} {\bibinfo {author} {\bibfnamefont {D.~L.}\ \bibnamefont
  {Moehring}}, \bibinfo {author} {\bibfnamefont {P.}~\bibnamefont {Maunz}},
  \bibinfo {author} {\bibfnamefont {S.}~\bibnamefont {Olmschenk}}, \bibinfo
  {author} {\bibfnamefont {K.~C.}\ \bibnamefont {Younge}}, \bibinfo {author}
  {\bibfnamefont {D.~N.}\ \bibnamefont {Matsukevich}}, \bibinfo {author}
  {\bibfnamefont {L.-M.}\ \bibnamefont {Duan}}, \ and\ \bibinfo {author}
  {\bibfnamefont {C.}~\bibnamefont {Monroe}},\ }\href {\doibase
  10.1038/nature06118} {\bibfield  {journal} {\bibinfo  {journal} {Nature}\
  }\textbf {\bibinfo {volume} {449}},\ \bibinfo {pages} {68} (\bibinfo {year}
  {2007})}\BibitemShut {NoStop}%
\bibitem [{\citenamefont {Hucul}\ \emph {et~al.}(2014)\citenamefont {Hucul},
  \citenamefont {Inlek}, \citenamefont {Vittorini}, \citenamefont {Crocker},
  \citenamefont {Debnath}, \citenamefont {Clark},\ and\ \citenamefont
  {Monroe}}]{hucul_modular_2014}%
  \BibitemOpen
  \bibfield  {author} {\bibinfo {author} {\bibfnamefont {D.}~\bibnamefont
  {Hucul}}, \bibinfo {author} {\bibfnamefont {I.~V.}\ \bibnamefont {Inlek}},
  \bibinfo {author} {\bibfnamefont {G.}~\bibnamefont {Vittorini}}, \bibinfo
  {author} {\bibfnamefont {C.}~\bibnamefont {Crocker}}, \bibinfo {author}
  {\bibfnamefont {S.}~\bibnamefont {Debnath}}, \bibinfo {author} {\bibfnamefont
  {S.~M.}\ \bibnamefont {Clark}}, \ and\ \bibinfo {author} {\bibfnamefont
  {C.}~\bibnamefont {Monroe}},\ }\href {http://arxiv.org/abs/1403.3696}
  {\bibfield  {journal} {\bibinfo  {journal} {{arXiv}:1403.3696 [quant-ph]}\ }
  (\bibinfo {year} {2014})}\BibitemShut {NoStop}%
\bibitem [{\citenamefont {Weitenberg}\ \emph {et~al.}(2011)\citenamefont
  {Weitenberg}, \citenamefont {Kuhr}, \citenamefont {Mølmer},\ and\
  \citenamefont {Sherson}}]{weitenberg_quantum_2011}%
  \BibitemOpen
  \bibfield  {author} {\bibinfo {author} {\bibfnamefont {C.}~\bibnamefont
  {Weitenberg}}, \bibinfo {author} {\bibfnamefont {S.}~\bibnamefont {Kuhr}},
  \bibinfo {author} {\bibfnamefont {K.}~\bibnamefont {Mølmer}}, \ and\
  \bibinfo {author} {\bibfnamefont {J.~F.}\ \bibnamefont {Sherson}},\ }\href
  {\doibase 10.1103/PhysRevA.84.032322} {\bibfield  {journal} {\bibinfo
  {journal} {Physical Review A}\ }\textbf {\bibinfo {volume} {84}},\ \bibinfo
  {pages} {032322} (\bibinfo {year} {2011})}\BibitemShut {NoStop}%
\bibitem [{\citenamefont {Yao}\ \emph {et~al.}(2011)\citenamefont {Yao},
  \citenamefont {Jiang}, \citenamefont {Gorshkov}, \citenamefont {Gong},
  \citenamefont {Zhai}, \citenamefont {Duan},\ and\ \citenamefont
  {Lukin}}]{yao_robust_2011}%
  \BibitemOpen
  \bibfield  {author} {\bibinfo {author} {\bibfnamefont {N.}~\bibnamefont
  {Yao}}, \bibinfo {author} {\bibfnamefont {L.}~\bibnamefont {Jiang}}, \bibinfo
  {author} {\bibfnamefont {A.}~\bibnamefont {Gorshkov}}, \bibinfo {author}
  {\bibfnamefont {Z.-X.}\ \bibnamefont {Gong}}, \bibinfo {author}
  {\bibfnamefont {A.}~\bibnamefont {Zhai}}, \bibinfo {author} {\bibfnamefont
  {L.-M.}\ \bibnamefont {Duan}}, \ and\ \bibinfo {author} {\bibfnamefont
  {M.}~\bibnamefont {Lukin}},\ }\href {\doibase 10.1103/PhysRevLett.106.040505}
  {\bibfield  {journal} {\bibinfo  {journal} {Physical Review Letters}\
  }\textbf {\bibinfo {volume} {106}} (\bibinfo {year} {2011}),\
  10.1103/PhysRevLett.106.040505}\BibitemShut {NoStop}%
\bibitem [{\citenamefont {Banchi}\ \emph
  {et~al.}(2011{\natexlab{a}})\citenamefont {Banchi}, \citenamefont {Bayat},
  \citenamefont {Verrucchi},\ and\ \citenamefont
  {Bose}}]{banchi_nonperturbative_2011}%
  \BibitemOpen
  \bibfield  {author} {\bibinfo {author} {\bibfnamefont {L.}~\bibnamefont
  {Banchi}}, \bibinfo {author} {\bibfnamefont {A.}~\bibnamefont {Bayat}},
  \bibinfo {author} {\bibfnamefont {P.}~\bibnamefont {Verrucchi}}, \ and\
  \bibinfo {author} {\bibfnamefont {S.}~\bibnamefont {Bose}},\ }\href {\doibase
  10.1103/PhysRevLett.106.140501} {\bibfield  {journal} {\bibinfo  {journal}
  {Physical Review Letters}\ }\textbf {\bibinfo {volume} {106}} (\bibinfo
  {year} {2011}{\natexlab{a}}),\ 10.1103/PhysRevLett.106.140501}\BibitemShut
  {NoStop}%
\bibitem [{\citenamefont {Ping}\ \emph {et~al.}(2013)\citenamefont {Ping},
  \citenamefont {Lovett}, \citenamefont {Benjamin},\ and\ \citenamefont
  {Gauger}}]{ping_practicality_2013}%
  \BibitemOpen
  \bibfield  {author} {\bibinfo {author} {\bibfnamefont {Y.}~\bibnamefont
  {Ping}}, \bibinfo {author} {\bibfnamefont {B.~W.}\ \bibnamefont {Lovett}},
  \bibinfo {author} {\bibfnamefont {S.~C.}\ \bibnamefont {Benjamin}}, \ and\
  \bibinfo {author} {\bibfnamefont {E.~M.}\ \bibnamefont {Gauger}},\
  }\href@noop {} {\bibfield  {journal} {\bibinfo  {journal} {Physical review
  letters}\ }\textbf {\bibinfo {volume} {110}},\ \bibinfo {pages} {100503}
  (\bibinfo {year} {2013})}\BibitemShut {NoStop}%
\bibitem [{\citenamefont {Calabrese}\ and\ \citenamefont
  {Cardy}(2005)}]{calabrese_evolution_2005}%
  \BibitemOpen
  \bibfield  {author} {\bibinfo {author} {\bibfnamefont {P.}~\bibnamefont
  {Calabrese}}\ and\ \bibinfo {author} {\bibfnamefont {J.}~\bibnamefont
  {Cardy}},\ }\href {\doibase 10.1088/1742-5468/2005/04/P04010} {\bibfield
  {journal} {\bibinfo  {journal} {Journal of Statistical Mechanics: Theory and
  Experiment}\ }\textbf {\bibinfo {volume} {2005}},\ \bibinfo {pages} {P04010}
  (\bibinfo {year} {2005})}\BibitemShut {NoStop}%
\bibitem [{\citenamefont {De~Chiara}\ \emph {et~al.}(2006)\citenamefont
  {De~Chiara}, \citenamefont {Montangero}, \citenamefont {Calabrese},\ and\
  \citenamefont {Fazio}}]{de_chiara_entanglement_2006}%
  \BibitemOpen
  \bibfield  {author} {\bibinfo {author} {\bibfnamefont {G.}~\bibnamefont
  {De~Chiara}}, \bibinfo {author} {\bibfnamefont {S.}~\bibnamefont
  {Montangero}}, \bibinfo {author} {\bibfnamefont {P.}~\bibnamefont
  {Calabrese}}, \ and\ \bibinfo {author} {\bibfnamefont {R.}~\bibnamefont
  {Fazio}},\ }\href {\doibase 10.1088/1742-5468/2006/03/P03001} {\bibfield
  {journal} {\bibinfo  {journal} {Journal of Statistical Mechanics: Theory and
  Experiment}\ }\textbf {\bibinfo {volume} {2006}},\ \bibinfo {pages} {P03001}
  (\bibinfo {year} {2006})}\BibitemShut {NoStop}%
\bibitem [{\citenamefont {Das}\ \emph {et~al.}(2011)\citenamefont {Das},
  \citenamefont {Garnerone},\ and\ \citenamefont
  {Haas}}]{das_entanglement_2011}%
  \BibitemOpen
  \bibfield  {author} {\bibinfo {author} {\bibfnamefont {A.}~\bibnamefont
  {Das}}, \bibinfo {author} {\bibfnamefont {S.}~\bibnamefont {Garnerone}}, \
  and\ \bibinfo {author} {\bibfnamefont {S.}~\bibnamefont {Haas}},\ }\href
  {\doibase 10.1103/PhysRevA.84.052317} {\bibfield  {journal} {\bibinfo
  {journal} {Physical Review A}\ }\textbf {\bibinfo {volume} {84}},\ \bibinfo
  {pages} {052317} (\bibinfo {year} {2011})}\BibitemShut {NoStop}%
\bibitem [{\citenamefont {Barmettler}\ \emph {et~al.}(2008)\citenamefont
  {Barmettler}, \citenamefont {Rey}, \citenamefont {Demler}, \citenamefont
  {Lukin}, \citenamefont {Bloch},\ and\ \citenamefont
  {Gritsev}}]{barmettler_quantum_2008}%
  \BibitemOpen
  \bibfield  {author} {\bibinfo {author} {\bibfnamefont {P.}~\bibnamefont
  {Barmettler}}, \bibinfo {author} {\bibfnamefont {A.}~\bibnamefont {Rey}},
  \bibinfo {author} {\bibfnamefont {E.}~\bibnamefont {Demler}}, \bibinfo
  {author} {\bibfnamefont {M.}~\bibnamefont {Lukin}}, \bibinfo {author}
  {\bibfnamefont {I.}~\bibnamefont {Bloch}}, \ and\ \bibinfo {author}
  {\bibfnamefont {V.}~\bibnamefont {Gritsev}},\ }\href {\doibase
  10.1103/PhysRevA.78.012330} {\bibfield  {journal} {\bibinfo  {journal} {Phys.
  Rev. A}\ }\textbf {\bibinfo {volume} {78}},\ \bibinfo {pages} {012330}
  (\bibinfo {year} {2008})}\BibitemShut {NoStop}%
\bibitem [{\citenamefont {Sengupta}\ and\ \citenamefont
  {Sen}(2009)}]{sengupta_entanglement_2009}%
  \BibitemOpen
  \bibfield  {author} {\bibinfo {author} {\bibfnamefont {K.}~\bibnamefont
  {Sengupta}}\ and\ \bibinfo {author} {\bibfnamefont {D.}~\bibnamefont {Sen}},\
  }\href {\doibase 10.1103/PhysRevA.80.032304} {\bibfield  {journal} {\bibinfo
  {journal} {Physical Review A}\ }\textbf {\bibinfo {volume} {80}},\ \bibinfo
  {pages} {032304} (\bibinfo {year} {2009})}\BibitemShut {NoStop}%
\bibitem [{\citenamefont {Sadiek}\ \emph {et~al.}(2010)\citenamefont {Sadiek},
  \citenamefont {Alkurtass},\ and\ \citenamefont
  {Aldossary}}]{sadiek_entanglement_2010}%
  \BibitemOpen
  \bibfield  {author} {\bibinfo {author} {\bibfnamefont {G.}~\bibnamefont
  {Sadiek}}, \bibinfo {author} {\bibfnamefont {B.}~\bibnamefont {Alkurtass}}, \
  and\ \bibinfo {author} {\bibfnamefont {O.}~\bibnamefont {Aldossary}},\ }\href
  {\doibase 10.1103/PhysRevA.82.052337} {\bibfield  {journal} {\bibinfo
  {journal} {Physical Review A}\ }\textbf {\bibinfo {volume} {82}},\ \bibinfo
  {pages} {052337} (\bibinfo {year} {2010})}\BibitemShut {NoStop}%
\bibitem [{\citenamefont {Alkurtass}\ \emph {et~al.}(2011)\citenamefont
  {Alkurtass}, \citenamefont {Sadiek},\ and\ \citenamefont
  {Kais}}]{alkurtass_entanglement_2011}%
  \BibitemOpen
  \bibfield  {author} {\bibinfo {author} {\bibfnamefont {B.}~\bibnamefont
  {Alkurtass}}, \bibinfo {author} {\bibfnamefont {G.}~\bibnamefont {Sadiek}}, \
  and\ \bibinfo {author} {\bibfnamefont {S.}~\bibnamefont {Kais}},\ }\href
  {\doibase 10.1103/PhysRevA.84.022314} {\bibfield  {journal} {\bibinfo
  {journal} {Physical Review A}\ }\textbf {\bibinfo {volume} {84}},\ \bibinfo
  {pages} {022314} (\bibinfo {year} {2011})}\BibitemShut {NoStop}%
\bibitem [{\citenamefont {Sadiek}\ \emph {et~al.}(2013)\citenamefont {Sadiek},
  \citenamefont {Xu},\ and\ \citenamefont {Kais}}]{sadiek_dynamics_2013}%
  \BibitemOpen
  \bibfield  {author} {\bibinfo {author} {\bibfnamefont {G.}~\bibnamefont
  {Sadiek}}, \bibinfo {author} {\bibfnamefont {Q.}~\bibnamefont {Xu}}, \ and\
  \bibinfo {author} {\bibfnamefont {S.}~\bibnamefont {Kais}},\ }\href@noop {}
  {\bibfield  {journal} {\bibinfo  {journal} {{arXiv} preprint
  {arXiv}:1304.5569}\ } (\bibinfo {year} {2013})}\BibitemShut {NoStop}%
\bibitem [{\citenamefont {Eisert}\ \emph {et~al.}(2004)\citenamefont {Eisert},
  \citenamefont {Plenio}, \citenamefont {Bose},\ and\ \citenamefont
  {Hartley}}]{eisert_towards_2004}%
  \BibitemOpen
  \bibfield  {author} {\bibinfo {author} {\bibfnamefont {J.}~\bibnamefont
  {Eisert}}, \bibinfo {author} {\bibfnamefont {M.~B.}\ \bibnamefont {Plenio}},
  \bibinfo {author} {\bibfnamefont {S.}~\bibnamefont {Bose}}, \ and\ \bibinfo
  {author} {\bibfnamefont {J.}~\bibnamefont {Hartley}},\ }\href@noop {}
  {\bibfield  {journal} {\bibinfo  {journal} {Physical review letters}\
  }\textbf {\bibinfo {volume} {93}},\ \bibinfo {pages} {190402} (\bibinfo
  {year} {2004})}\BibitemShut {NoStop}%
\bibitem [{\citenamefont {Schachenmayer}\ \emph {et~al.}(2013)\citenamefont
  {Schachenmayer}, \citenamefont {Lanyon}, \citenamefont {Roos},\ and\
  \citenamefont {Daley}}]{schachenmayer_entanglement_2013}%
  \BibitemOpen
  \bibfield  {author} {\bibinfo {author} {\bibfnamefont {J.}~\bibnamefont
  {Schachenmayer}}, \bibinfo {author} {\bibfnamefont {B.}~\bibnamefont
  {Lanyon}}, \bibinfo {author} {\bibfnamefont {C.}~\bibnamefont {Roos}}, \ and\
  \bibinfo {author} {\bibfnamefont {A.}~\bibnamefont {Daley}},\ }\href@noop {}
  {\bibfield  {journal} {\bibinfo  {journal} {{arXiv} preprint
  {arXiv}:1305.6880}\ } (\bibinfo {year} {2013})}\BibitemShut {NoStop}%
\bibitem [{\citenamefont {Hauke}\ and\ \citenamefont
  {Tagliacozzo}(2013)}]{hauke_spread_2013}%
  \BibitemOpen
  \bibfield  {author} {\bibinfo {author} {\bibfnamefont {P.}~\bibnamefont
  {Hauke}}\ and\ \bibinfo {author} {\bibfnamefont {L.}~\bibnamefont
  {Tagliacozzo}},\ }\href {\doibase 10.1103/PhysRevLett.111.207202} {\bibfield
  {journal} {\bibinfo  {journal} {Physical Review Letters}\ }\textbf {\bibinfo
  {volume} {111}},\ \bibinfo {pages} {207202} (\bibinfo {year}
  {2013})}\BibitemShut {NoStop}%
\bibitem [{\citenamefont {Kim}\ and\ \citenamefont
  {Huse}(2013)}]{kim_ballistic_2013}%
  \BibitemOpen
  \bibfield  {author} {\bibinfo {author} {\bibfnamefont {H.}~\bibnamefont
  {Kim}}\ and\ \bibinfo {author} {\bibfnamefont {D.~A.}\ \bibnamefont {Huse}},\
  }\href {\doibase 10.1103/PhysRevLett.111.127205} {\bibfield  {journal}
  {\bibinfo  {journal} {Physical Review Letters}\ }\textbf {\bibinfo {volume}
  {111}},\ \bibinfo {pages} {127205} (\bibinfo {year} {2013})}\BibitemShut
  {NoStop}%
\bibitem [{\citenamefont {Bardarson}\ \emph {et~al.}(2012)\citenamefont
  {Bardarson}, \citenamefont {Pollmann},\ and\ \citenamefont
  {Moore}}]{bardarson_unbounded_2012}%
  \BibitemOpen
  \bibfield  {author} {\bibinfo {author} {\bibfnamefont {J.~H.}\ \bibnamefont
  {Bardarson}}, \bibinfo {author} {\bibfnamefont {F.}~\bibnamefont {Pollmann}},
  \ and\ \bibinfo {author} {\bibfnamefont {J.~E.}\ \bibnamefont {Moore}},\
  }\href {\doibase 10.1103/PhysRevLett.109.017202} {\bibfield  {journal}
  {\bibinfo  {journal} {Physical Review Letters}\ }\textbf {\bibinfo {volume}
  {109}},\ \bibinfo {pages} {017202} (\bibinfo {year} {2012})}\BibitemShut
  {NoStop}%
\bibitem [{\citenamefont {Serbyn}\ \emph {et~al.}(2013)\citenamefont {Serbyn},
  \citenamefont {Papić},\ and\ \citenamefont
  {Abanin}}]{serbyn_universal_2013}%
  \BibitemOpen
  \bibfield  {author} {\bibinfo {author} {\bibfnamefont {M.}~\bibnamefont
  {Serbyn}}, \bibinfo {author} {\bibfnamefont {Z.}~\bibnamefont {Papić}}, \
  and\ \bibinfo {author} {\bibfnamefont {D.~A.}\ \bibnamefont {Abanin}},\
  }\href {\doibase 10.1103/PhysRevLett.110.260601} {\bibfield  {journal}
  {\bibinfo  {journal} {Physical Review Letters}\ }\textbf {\bibinfo {volume}
  {110}},\ \bibinfo {pages} {260601} (\bibinfo {year} {2013})}\BibitemShut
  {NoStop}%
\bibitem [{\citenamefont {Vosk}\ and\ \citenamefont
  {Altman}(2013)}]{vosk_many-body_2013}%
  \BibitemOpen
  \bibfield  {author} {\bibinfo {author} {\bibfnamefont {R.}~\bibnamefont
  {Vosk}}\ and\ \bibinfo {author} {\bibfnamefont {E.}~\bibnamefont {Altman}},\
  }\href {\doibase 10.1103/PhysRevLett.110.067204} {\bibfield  {journal}
  {\bibinfo  {journal} {Physical Review Letters}\ }\textbf {\bibinfo {volume}
  {110}},\ \bibinfo {pages} {067204} (\bibinfo {year} {2013})}\BibitemShut
  {NoStop}%
\bibitem [{\citenamefont {Jurcevic}\ \emph {et~al.}(2014)\citenamefont
  {Jurcevic}, \citenamefont {Lanyon}, \citenamefont {Hauke}, \citenamefont
  {Hempel}, \citenamefont {Zoller}, \citenamefont {Blatt},\ and\ \citenamefont
  {Roos}}]{jurcevic_observation_2014}%
  \BibitemOpen
  \bibfield  {author} {\bibinfo {author} {\bibfnamefont {P.}~\bibnamefont
  {Jurcevic}}, \bibinfo {author} {\bibfnamefont {B.~P.}\ \bibnamefont
  {Lanyon}}, \bibinfo {author} {\bibfnamefont {P.}~\bibnamefont {Hauke}},
  \bibinfo {author} {\bibfnamefont {C.}~\bibnamefont {Hempel}}, \bibinfo
  {author} {\bibfnamefont {P.}~\bibnamefont {Zoller}}, \bibinfo {author}
  {\bibfnamefont {R.}~\bibnamefont {Blatt}}, \ and\ \bibinfo {author}
  {\bibfnamefont {C.~F.}\ \bibnamefont {Roos}},\ }\href
  {http://arxiv.org/abs/1401.5387} {\bibfield  {journal} {\bibinfo  {journal}
  {{arXiv}:1401.5387 [cond-mat, physics:physics, physics:quant-ph]}\ }
  (\bibinfo {year} {2014})}\BibitemShut {NoStop}%
\bibitem [{\citenamefont {Richerme}\ \emph {et~al.}(2014)\citenamefont
  {Richerme}, \citenamefont {Gong}, \citenamefont {Lee}, \citenamefont {Senko},
  \citenamefont {Smith}, \citenamefont {Foss-Feig}, \citenamefont {Michalakis},
  \citenamefont {Gorshkov},\ and\ \citenamefont
  {Monroe}}]{richerme_non-local_2014}%
  \BibitemOpen
  \bibfield  {author} {\bibinfo {author} {\bibfnamefont {P.}~\bibnamefont
  {Richerme}}, \bibinfo {author} {\bibfnamefont {Z.-X.}\ \bibnamefont {Gong}},
  \bibinfo {author} {\bibfnamefont {A.}~\bibnamefont {Lee}}, \bibinfo {author}
  {\bibfnamefont {C.}~\bibnamefont {Senko}}, \bibinfo {author} {\bibfnamefont
  {J.}~\bibnamefont {Smith}}, \bibinfo {author} {\bibfnamefont
  {M.}~\bibnamefont {Foss-Feig}}, \bibinfo {author} {\bibfnamefont
  {S.}~\bibnamefont {Michalakis}}, \bibinfo {author} {\bibfnamefont {A.~V.}\
  \bibnamefont {Gorshkov}}, \ and\ \bibinfo {author} {\bibfnamefont
  {C.}~\bibnamefont {Monroe}},\ }\href {http://arxiv.org/abs/1401.5088}
  {\bibfield  {journal} {\bibinfo  {journal} {{arXiv}:1401.5088 [quant-ph]}\ }
  (\bibinfo {year} {2014})}\BibitemShut {NoStop}%
\bibitem [{\citenamefont {Fukuhara}\ \emph
  {et~al.}(2013{\natexlab{a}})\citenamefont {Fukuhara}, \citenamefont
  {Schauß}, \citenamefont {Endres}, \citenamefont {Hild}, \citenamefont
  {Cheneau}, \citenamefont {Bloch},\ and\ \citenamefont
  {Gross}}]{fukuhara_microscopic_2013}%
  \BibitemOpen
  \bibfield  {author} {\bibinfo {author} {\bibfnamefont {T.}~\bibnamefont
  {Fukuhara}}, \bibinfo {author} {\bibfnamefont {P.}~\bibnamefont {Schauß}},
  \bibinfo {author} {\bibfnamefont {M.}~\bibnamefont {Endres}}, \bibinfo
  {author} {\bibfnamefont {S.}~\bibnamefont {Hild}}, \bibinfo {author}
  {\bibfnamefont {M.}~\bibnamefont {Cheneau}}, \bibinfo {author} {\bibfnamefont
  {I.}~\bibnamefont {Bloch}}, \ and\ \bibinfo {author} {\bibfnamefont
  {C.}~\bibnamefont {Gross}},\ }\href {\doibase 10.1038/nature12541} {\bibfield
   {journal} {\bibinfo  {journal} {Nature}\ }\textbf {\bibinfo {volume}
  {502}},\ \bibinfo {pages} {76} (\bibinfo {year}
  {2013}{\natexlab{a}})}\BibitemShut {NoStop}%
\bibitem [{\citenamefont {Campos~Venuti}\ \emph {et~al.}(2006)\citenamefont
  {Campos~Venuti}, \citenamefont {Degli Esposti~Boschi},\ and\ \citenamefont
  {Roncaglia}}]{campos_venuti_long-distance_2006}%
  \BibitemOpen
  \bibfield  {author} {\bibinfo {author} {\bibfnamefont {L.}~\bibnamefont
  {Campos~Venuti}}, \bibinfo {author} {\bibfnamefont {C.}~\bibnamefont {Degli
  Esposti~Boschi}}, \ and\ \bibinfo {author} {\bibfnamefont {M.}~\bibnamefont
  {Roncaglia}},\ }\href@noop {} {\bibfield  {journal} {\bibinfo  {journal}
  {Physical review letters}\ }\textbf {\bibinfo {volume} {96}},\ \bibinfo
  {pages} {247206} (\bibinfo {year} {2006})}\BibitemShut {NoStop}%
\bibitem [{\citenamefont {Campos~Venuti}\ \emph {et~al.}(2007)\citenamefont
  {Campos~Venuti}, \citenamefont {Giampaolo}, \citenamefont {Illuminati},
  \citenamefont {Zanardi},\ and\ \citenamefont
  {Campos~Venuti}}]{campos_venuti_long-distance_2007}%
  \BibitemOpen
  \bibfield  {author} {\bibinfo {author} {\bibfnamefont {L.}~\bibnamefont
  {Campos~Venuti}}, \bibinfo {author} {\bibfnamefont {S.~M.}\ \bibnamefont
  {Giampaolo}}, \bibinfo {author} {\bibfnamefont {F.}~\bibnamefont
  {Illuminati}}, \bibinfo {author} {\bibfnamefont {P.}~\bibnamefont {Zanardi}},
  \ and\ \bibinfo {author} {\bibfnamefont {L.}~\bibnamefont {Campos~Venuti}},\
  }\href {\doibase 10.1103/PhysRevA.76.052328} {\bibfield  {journal} {\bibinfo
  {journal} {Physical Review A}\ }\textbf {\bibinfo {volume} {76}},\ \bibinfo
  {pages} {52328} (\bibinfo {year} {2007})}\BibitemShut {NoStop}%
\bibitem [{\citenamefont {Giampaolo}\ and\ \citenamefont
  {Illuminati}(2010)}]{giampaolo_long-distance_2010}%
  \BibitemOpen
  \bibfield  {author} {\bibinfo {author} {\bibfnamefont {S.~M.}\ \bibnamefont
  {Giampaolo}}\ and\ \bibinfo {author} {\bibfnamefont {F.}~\bibnamefont
  {Illuminati}},\ }\href {http://stacks.iop.org/1367-2630/12/i=2/a=025019}
  {\bibfield  {journal} {\bibinfo  {journal} {New Journal of Physics}\ }\textbf
  {\bibinfo {volume} {12}},\ \bibinfo {pages} {25019} (\bibinfo {year}
  {2010})}\BibitemShut {NoStop}%
\bibitem [{\citenamefont {Zippilli}\ \emph {et~al.}(2013)\citenamefont
  {Zippilli}, \citenamefont {Johanning}, \citenamefont {Giampaolo},
  \citenamefont {Wunderlich},\ and\ \citenamefont
  {Illuminati}}]{zippilli_adiabatic_2013}%
  \BibitemOpen
  \bibfield  {author} {\bibinfo {author} {\bibfnamefont {S.}~\bibnamefont
  {Zippilli}}, \bibinfo {author} {\bibfnamefont {M.}~\bibnamefont {Johanning}},
  \bibinfo {author} {\bibfnamefont {S.~M.}\ \bibnamefont {Giampaolo}}, \bibinfo
  {author} {\bibfnamefont {C.}~\bibnamefont {Wunderlich}}, \ and\ \bibinfo
  {author} {\bibfnamefont {F.}~\bibnamefont {Illuminati}},\ }\href@noop {}
  {\bibfield  {journal} {\bibinfo  {journal} {{arXiv} preprint
  {arXiv}:1304.0261}\ } (\bibinfo {year} {2013})}\BibitemShut {NoStop}%
\bibitem [{\citenamefont {Wichterich}\ and\ \citenamefont
  {Bose}(2009)}]{wichterich_exploiting_2009}%
  \BibitemOpen
  \bibfield  {author} {\bibinfo {author} {\bibfnamefont {H.}~\bibnamefont
  {Wichterich}}\ and\ \bibinfo {author} {\bibfnamefont {S.}~\bibnamefont
  {Bose}},\ }\href {\doibase 10.1103/PhysRevA.79.060302} {\bibfield  {journal}
  {\bibinfo  {journal} {Phys. Rev. A}\ }\textbf {\bibinfo {volume} {79}},\
  \bibinfo {pages} {60302} (\bibinfo {year} {2009})}\BibitemShut {NoStop}%
\bibitem [{\citenamefont {Alkurtass}\ \emph {et~al.}(2013)\citenamefont
  {Alkurtass}, \citenamefont {Wichterich},\ and\ \citenamefont
  {Bose}}]{alkurtass_quench_2013}%
  \BibitemOpen
  \bibfield  {author} {\bibinfo {author} {\bibfnamefont {B.}~\bibnamefont
  {Alkurtass}}, \bibinfo {author} {\bibfnamefont {H.}~\bibnamefont
  {Wichterich}}, \ and\ \bibinfo {author} {\bibfnamefont {S.}~\bibnamefont
  {Bose}},\ }\href@noop {} {\bibfield  {journal} {\bibinfo  {journal} {{arXiv}
  preprint {arXiv}:1309.5756}\ } (\bibinfo {year} {2013})}\BibitemShut
  {NoStop}%
\bibitem [{\citenamefont {Bayat}\ \emph {et~al.}(2010)\citenamefont {Bayat},
  \citenamefont {Bose},\ and\ \citenamefont
  {Sodano}}]{bayat_entanglement_2010}%
  \BibitemOpen
  \bibfield  {author} {\bibinfo {author} {\bibfnamefont {A.}~\bibnamefont
  {Bayat}}, \bibinfo {author} {\bibfnamefont {S.}~\bibnamefont {Bose}}, \ and\
  \bibinfo {author} {\bibfnamefont {P.}~\bibnamefont {Sodano}},\ }\href
  {\doibase 10.1103/PhysRevLett.105.187204} {\bibfield  {journal} {\bibinfo
  {journal} {Physical Review Letters}\ }\textbf {\bibinfo {volume} {105}},\
  \bibinfo {pages} {187204} (\bibinfo {year} {2010})}\BibitemShut {NoStop}%
\bibitem [{\citenamefont {Sodano}\ \emph {et~al.}(2010)\citenamefont {Sodano},
  \citenamefont {Bayat},\ and\ \citenamefont {Bose}}]{sodano_kondo_2010}%
  \BibitemOpen
  \bibfield  {author} {\bibinfo {author} {\bibfnamefont {P.}~\bibnamefont
  {Sodano}}, \bibinfo {author} {\bibfnamefont {A.}~\bibnamefont {Bayat}}, \
  and\ \bibinfo {author} {\bibfnamefont {S.}~\bibnamefont {Bose}},\ }\href
  {\doibase 10.1103/PhysRevB.81.100412} {\bibfield  {journal} {\bibinfo
  {journal} {Physical Review B}\ }\textbf {\bibinfo {volume} {81}},\ \bibinfo
  {pages} {100412} (\bibinfo {year} {2010})}\BibitemShut {NoStop}%
\bibitem [{\citenamefont {Bayat}\ \emph {et~al.}(2011)\citenamefont {Bayat},
  \citenamefont {Banchi}, \citenamefont {Bose},\ and\ \citenamefont
  {Verrucchi}}]{bayat_initializing_2011}%
  \BibitemOpen
  \bibfield  {author} {\bibinfo {author} {\bibfnamefont {A.}~\bibnamefont
  {Bayat}}, \bibinfo {author} {\bibfnamefont {L.}~\bibnamefont {Banchi}},
  \bibinfo {author} {\bibfnamefont {S.}~\bibnamefont {Bose}}, \ and\ \bibinfo
  {author} {\bibfnamefont {P.}~\bibnamefont {Verrucchi}},\ }\href {\doibase
  10.1103/PhysRevA.83.062328} {\bibfield  {journal} {\bibinfo  {journal}
  {Physical Review A}\ }\textbf {\bibinfo {volume} {83}} (\bibinfo {year}
  {2011}),\ 10.1103/PhysRevA.83.062328}\BibitemShut {NoStop}%
\bibitem [{\citenamefont {Sansoni}\ \emph {et~al.}(2012)\citenamefont
  {Sansoni}, \citenamefont {Sciarrino}, \citenamefont {Vallone}, \citenamefont
  {Mataloni}, \citenamefont {Crespi}, \citenamefont {Ramponi},\ and\
  \citenamefont {Osellame}}]{sansoni_two-particle_2012}%
  \BibitemOpen
  \bibfield  {author} {\bibinfo {author} {\bibfnamefont {L.}~\bibnamefont
  {Sansoni}}, \bibinfo {author} {\bibfnamefont {F.}~\bibnamefont {Sciarrino}},
  \bibinfo {author} {\bibfnamefont {G.}~\bibnamefont {Vallone}}, \bibinfo
  {author} {\bibfnamefont {P.}~\bibnamefont {Mataloni}}, \bibinfo {author}
  {\bibfnamefont {A.}~\bibnamefont {Crespi}}, \bibinfo {author} {\bibfnamefont
  {R.}~\bibnamefont {Ramponi}}, \ and\ \bibinfo {author} {\bibfnamefont
  {R.}~\bibnamefont {Osellame}},\ }\href {\doibase
  10.1103/PhysRevLett.108.010502} {\bibfield  {journal} {\bibinfo  {journal}
  {Physical Review Letters}\ }\textbf {\bibinfo {volume} {108}},\ \bibinfo
  {pages} {010502} (\bibinfo {year} {2012})}\BibitemShut {NoStop}%
\bibitem [{\citenamefont {Aaronson}\ and\ \citenamefont
  {Arkhipov}(2011)}]{aaronson_computational_2011}%
  \BibitemOpen
  \bibfield  {author} {\bibinfo {author} {\bibfnamefont {S.}~\bibnamefont
  {Aaronson}}\ and\ \bibinfo {author} {\bibfnamefont {A.}~\bibnamefont
  {Arkhipov}},\ }in\ \href {\doibase 10.1145/1993636.1993682} {\emph {\bibinfo
  {booktitle} {Proceedings of the Forty-third Annual {ACM} Symposium on Theory
  of Computing}}},\ \bibinfo {series and number} {{STOC} '11}\ (\bibinfo
  {publisher} {{ACM}},\ \bibinfo {address} {New York, {NY}, {USA}},\ \bibinfo
  {year} {2011})\ p.\ \bibinfo {pages} {333–342}\BibitemShut {NoStop}%
\bibitem [{\citenamefont {Terhal}\ and\ \citenamefont
  {DiVincenzo}(2002)}]{terhal_classical_2002}%
  \BibitemOpen
  \bibfield  {author} {\bibinfo {author} {\bibfnamefont {B.~M.}\ \bibnamefont
  {Terhal}}\ and\ \bibinfo {author} {\bibfnamefont {D.~P.}\ \bibnamefont
  {DiVincenzo}},\ }\href {\doibase 10.1103/PhysRevA.65.032325} {\bibfield
  {journal} {\bibinfo  {journal} {Physical Review A}\ }\textbf {\bibinfo
  {volume} {65}},\ \bibinfo {pages} {032325} (\bibinfo {year}
  {2002})}\BibitemShut {NoStop}%
\bibitem [{\citenamefont {Cheneau}\ \emph {et~al.}(2012)\citenamefont
  {Cheneau}, \citenamefont {Barmettler}, \citenamefont {Poletti}, \citenamefont
  {Endres}, \citenamefont {Schauß}, \citenamefont {Fukuhara}, \citenamefont
  {Gross}, \citenamefont {Bloch}, \citenamefont {Kollath},\ and\ \citenamefont
  {Kuhr}}]{cheneau_light-cone-like_2012}%
  \BibitemOpen
  \bibfield  {author} {\bibinfo {author} {\bibfnamefont {M.}~\bibnamefont
  {Cheneau}}, \bibinfo {author} {\bibfnamefont {P.}~\bibnamefont {Barmettler}},
  \bibinfo {author} {\bibfnamefont {D.}~\bibnamefont {Poletti}}, \bibinfo
  {author} {\bibfnamefont {M.}~\bibnamefont {Endres}}, \bibinfo {author}
  {\bibfnamefont {P.}~\bibnamefont {Schauß}}, \bibinfo {author} {\bibfnamefont
  {T.}~\bibnamefont {Fukuhara}}, \bibinfo {author} {\bibfnamefont
  {C.}~\bibnamefont {Gross}}, \bibinfo {author} {\bibfnamefont
  {I.}~\bibnamefont {Bloch}}, \bibinfo {author} {\bibfnamefont
  {C.}~\bibnamefont {Kollath}}, \ and\ \bibinfo {author} {\bibfnamefont
  {S.}~\bibnamefont {Kuhr}},\ }\href {\doibase 10.1038/nature10748} {\bibfield
  {journal} {\bibinfo  {journal} {Nature}\ }\textbf {\bibinfo {volume} {481}},\
  \bibinfo {pages} {484} (\bibinfo {year} {2012})}\BibitemShut {NoStop}%
\bibitem [{\citenamefont {Bose}(2003)}]{bose_quantum_2003}%
  \BibitemOpen
  \bibfield  {author} {\bibinfo {author} {\bibfnamefont {S.}~\bibnamefont
  {Bose}},\ }\href@noop {} {\bibfield  {journal} {\bibinfo  {journal} {Physical
  review letters}\ }\textbf {\bibinfo {volume} {91}},\ \bibinfo {pages}
  {207901} (\bibinfo {year} {2003})}\BibitemShut {NoStop}%
\bibitem [{\citenamefont {Childs}(2009)}]{childs_universal_2009}%
  \BibitemOpen
  \bibfield  {author} {\bibinfo {author} {\bibfnamefont {A.~M.}\ \bibnamefont
  {Childs}},\ }\href {\doibase 10.1103/PhysRevLett.102.180501} {\bibfield
  {journal} {\bibinfo  {journal} {Physical Review Letters}\ }\textbf {\bibinfo
  {volume} {102}},\ \bibinfo {pages} {180501} (\bibinfo {year}
  {2009})}\BibitemShut {NoStop}%
\bibitem [{\citenamefont {Kay}(2010)}]{kay_perfect_2010}%
  \BibitemOpen
  \bibfield  {author} {\bibinfo {author} {\bibfnamefont {A.}~\bibnamefont
  {Kay}},\ }\href {\doibase 10.1142/S0219749910006514} {\bibfield  {journal}
  {\bibinfo  {journal} {International Journal of Quantum Information}\ }\textbf
  {\bibinfo {volume} {08}},\ \bibinfo {pages} {641} (\bibinfo {year}
  {2010})}\BibitemShut {NoStop}%
\bibitem [{\citenamefont {Di~Franco}\ \emph {et~al.}(2008)\citenamefont
  {Di~Franco}, \citenamefont {Paternostro},\ and\ \citenamefont
  {Kim}}]{di_franco_nested_2008}%
  \BibitemOpen
  \bibfield  {author} {\bibinfo {author} {\bibfnamefont {C.}~\bibnamefont
  {Di~Franco}}, \bibinfo {author} {\bibfnamefont {M.}~\bibnamefont
  {Paternostro}}, \ and\ \bibinfo {author} {\bibfnamefont {M.~S.}\ \bibnamefont
  {Kim}},\ }\href {\doibase 10.1103/PhysRevA.77.020303} {\bibfield  {journal}
  {\bibinfo  {journal} {Physical Review A}\ }\textbf {\bibinfo {volume} {77}},\
  \bibinfo {pages} {020303} (\bibinfo {year} {2008})}\BibitemShut {NoStop}%
\bibitem [{\citenamefont {Banchi}(2013)}]{banchi_ballistic_2013}%
  \BibitemOpen
  \bibfield  {author} {\bibinfo {author} {\bibfnamefont {L.}~\bibnamefont
  {Banchi}},\ }\href {\doibase 10.1140/epjp/i2013-13137-6} {\bibfield
  {journal} {\bibinfo  {journal} {The European Physical Journal Plus}\ }\textbf
  {\bibinfo {volume} {128}},\ \bibinfo {pages} {1} (\bibinfo {year}
  {2013})}\BibitemShut {NoStop}%
\bibitem [{\citenamefont {Banchi}\ \emph {et~al.}(2010)\citenamefont {Banchi},
  \citenamefont {Apollaro}, \citenamefont {Cuccoli}, \citenamefont {Vaia},\
  and\ \citenamefont {Verrucchi}}]{banchi_optimal_2010}%
  \BibitemOpen
  \bibfield  {author} {\bibinfo {author} {\bibfnamefont {L.}~\bibnamefont
  {Banchi}}, \bibinfo {author} {\bibfnamefont {T.}~\bibnamefont {Apollaro}},
  \bibinfo {author} {\bibfnamefont {A.}~\bibnamefont {Cuccoli}}, \bibinfo
  {author} {\bibfnamefont {R.}~\bibnamefont {Vaia}}, \ and\ \bibinfo {author}
  {\bibfnamefont {P.}~\bibnamefont {Verrucchi}},\ }\href {\doibase
  10.1103/PhysRevA.82.052321} {\bibfield  {journal} {\bibinfo  {journal}
  {Physical Review A}\ }\textbf {\bibinfo {volume} {82}} (\bibinfo {year}
  {2010}),\ 10.1103/PhysRevA.82.052321}\BibitemShut {NoStop}%
\bibitem [{\citenamefont {Banchi}\ \emph
  {et~al.}(2011{\natexlab{b}})\citenamefont {Banchi}, \citenamefont {Apollaro},
  \citenamefont {Cuccoli}, \citenamefont {Vaia},\ and\ \citenamefont
  {Verrucchi}}]{banchi_long_2011}%
  \BibitemOpen
  \bibfield  {author} {\bibinfo {author} {\bibfnamefont {L.}~\bibnamefont
  {Banchi}}, \bibinfo {author} {\bibfnamefont {T.~J.~G.}\ \bibnamefont
  {Apollaro}}, \bibinfo {author} {\bibfnamefont {A.}~\bibnamefont {Cuccoli}},
  \bibinfo {author} {\bibfnamefont {R.}~\bibnamefont {Vaia}}, \ and\ \bibinfo
  {author} {\bibfnamefont {P.}~\bibnamefont {Verrucchi}},\ }\href
  {http://stacks.iop.org/1367-2630/13/i=12/a=123006} {\bibfield  {journal}
  {\bibinfo  {journal} {New Journal of Physics}\ }\textbf {\bibinfo {volume}
  {13}},\ \bibinfo {pages} {123006} (\bibinfo {year}
  {2011}{\natexlab{b}})}\BibitemShut {NoStop}%
\bibitem [{\citenamefont {Bose}\ \emph {et~al.}(2014)\citenamefont {Bose},
  \citenamefont {Bayat}, \citenamefont {Sodano}, \citenamefont {Banchi},\ and\
  \citenamefont {Verrucchi}}]{bose_spin_2014}%
  \BibitemOpen
  \bibfield  {author} {\bibinfo {author} {\bibfnamefont {S.}~\bibnamefont
  {Bose}}, \bibinfo {author} {\bibfnamefont {A.}~\bibnamefont {Bayat}},
  \bibinfo {author} {\bibfnamefont {P.}~\bibnamefont {Sodano}}, \bibinfo
  {author} {\bibfnamefont {L.}~\bibnamefont {Banchi}}, \ and\ \bibinfo {author}
  {\bibfnamefont {P.}~\bibnamefont {Verrucchi}},\ }in\ \href
  {http://link.springer.com/chapter/10.1007/978-3-642-39937-4_1} {\emph
  {\bibinfo {booktitle} {Quantum State Transfer and Network Engineering}}},\
  \bibinfo {series and number} {Quantum Science and Technology},\ \bibinfo
  {editor} {edited by\ \bibinfo {editor} {\bibfnamefont {G.~M.}\ \bibnamefont
  {Nikolopoulos}}\ and\ \bibinfo {editor} {\bibfnamefont {I.}~\bibnamefont
  {Jex}}}\ (\bibinfo  {publisher} {Springer Berlin Heidelberg},\ \bibinfo
  {year} {2014})\ pp.\ \bibinfo {pages} {1--37}\BibitemShut {NoStop}%
\bibitem [{\citenamefont {Koetsier}\ \emph {et~al.}(2008)\citenamefont
  {Koetsier}, \citenamefont {Duine}, \citenamefont {Bloch},\ and\ \citenamefont
  {Stoof}}]{koetsier_achieving_2008}%
  \BibitemOpen
  \bibfield  {author} {\bibinfo {author} {\bibfnamefont {A.}~\bibnamefont
  {Koetsier}}, \bibinfo {author} {\bibfnamefont {R.}~\bibnamefont {Duine}},
  \bibinfo {author} {\bibfnamefont {I.}~\bibnamefont {Bloch}}, \ and\ \bibinfo
  {author} {\bibfnamefont {H.}~\bibnamefont {Stoof}},\ }\href {\doibase
  10.1103/PhysRevA.77.023623} {\bibfield  {journal} {\bibinfo  {journal} {Phys.
  Rev. A}\ }\textbf {\bibinfo {volume} {77}},\ \bibinfo {pages} {023623–+}
  (\bibinfo {year} {2008})}\BibitemShut {NoStop}%
\bibitem [{\citenamefont {Lieb}(1967)}]{lieb_residual_1967}%
  \BibitemOpen
  \bibfield  {author} {\bibinfo {author} {\bibfnamefont {E.~H.}\ \bibnamefont
  {Lieb}},\ }\href@noop {} {\bibfield  {journal} {\bibinfo  {journal} {Physical
  Review}\ }\textbf {\bibinfo {volume} {162}},\ \bibinfo {pages} {162–172}
  (\bibinfo {year} {1967})}\BibitemShut {NoStop}%
\bibitem [{\citenamefont {Christandl}\ \emph {et~al.}(2004)\citenamefont
  {Christandl}, \citenamefont {Datta}, \citenamefont {Ekert},\ and\
  \citenamefont {Landahl}}]{christandl_perfect_2004}%
  \BibitemOpen
  \bibfield  {author} {\bibinfo {author} {\bibfnamefont {M.}~\bibnamefont
  {Christandl}}, \bibinfo {author} {\bibfnamefont {N.}~\bibnamefont {Datta}},
  \bibinfo {author} {\bibfnamefont {A.}~\bibnamefont {Ekert}}, \ and\ \bibinfo
  {author} {\bibfnamefont {A.~J.}\ \bibnamefont {Landahl}},\ }\href@noop {}
  {\bibfield  {journal} {\bibinfo  {journal} {Physical review letters}\
  }\textbf {\bibinfo {volume} {92}},\ \bibinfo {pages} {187902} (\bibinfo
  {year} {2004})}\BibitemShut {NoStop}%
\bibitem [{\citenamefont {Kay}(2009)}]{kay_review_2009}%
  \BibitemOpen
  \bibfield  {author} {\bibinfo {author} {\bibfnamefont {A.}~\bibnamefont
  {Kay}},\ }\href@noop {} {\bibfield  {journal} {\bibinfo  {journal} {Arxiv
  preprint {arXiv}:0903.4274}\ } (\bibinfo {year} {2009})}\BibitemShut
  {NoStop}%
\bibitem [{\citenamefont {Bennett}\ \emph {et~al.}(1996)\citenamefont
  {Bennett}, \citenamefont {DiVincenzo}, \citenamefont {Smolin},\ and\
  \citenamefont {Wootters}}]{bennett_mixed-state_1996}%
  \BibitemOpen
  \bibfield  {author} {\bibinfo {author} {\bibfnamefont {C.~H.}\ \bibnamefont
  {Bennett}}, \bibinfo {author} {\bibfnamefont {D.~P.}\ \bibnamefont
  {DiVincenzo}}, \bibinfo {author} {\bibfnamefont {J.~A.}\ \bibnamefont
  {Smolin}}, \ and\ \bibinfo {author} {\bibfnamefont {W.~K.}\ \bibnamefont
  {Wootters}},\ }\href@noop {} {\bibfield  {journal} {\bibinfo  {journal}
  {Physical Review A}\ }\textbf {\bibinfo {volume} {54}},\ \bibinfo {pages}
  {3824–3851} (\bibinfo {year} {1996})}\BibitemShut {NoStop}%
\bibitem [{\citenamefont {Trotzky}\ \emph {et~al.}(2010)\citenamefont
  {Trotzky}, \citenamefont {Chen}, \citenamefont {Schnorrberger}, \citenamefont
  {Cheinet},\ and\ \citenamefont {Bloch}}]{trotzky_controlling_2010}%
  \BibitemOpen
  \bibfield  {author} {\bibinfo {author} {\bibfnamefont {S.}~\bibnamefont
  {Trotzky}}, \bibinfo {author} {\bibfnamefont {Y.-A.}\ \bibnamefont {Chen}},
  \bibinfo {author} {\bibfnamefont {U.}~\bibnamefont {Schnorrberger}}, \bibinfo
  {author} {\bibfnamefont {P.}~\bibnamefont {Cheinet}}, \ and\ \bibinfo
  {author} {\bibfnamefont {I.}~\bibnamefont {Bloch}},\ }\href@noop {} {\
  (\bibinfo {year} {2010})}\BibitemShut {NoStop}%
\bibitem [{\citenamefont {Negrevergne}\ \emph {et~al.}(2006)\citenamefont
  {Negrevergne}, \citenamefont {Mahesh}, \citenamefont {Ryan}, \citenamefont
  {Ditty}, \citenamefont {Cyr-Racine}, \citenamefont {Power}, \citenamefont
  {Boulant}, \citenamefont {Havel}, \citenamefont {Cory},\ and\ \citenamefont
  {Laflamme}}]{negrevergne_benchmarking_2006}%
  \BibitemOpen
  \bibfield  {author} {\bibinfo {author} {\bibfnamefont {C.}~\bibnamefont
  {Negrevergne}}, \bibinfo {author} {\bibfnamefont {T.~S.}\ \bibnamefont
  {Mahesh}}, \bibinfo {author} {\bibfnamefont {C.~A.}\ \bibnamefont {Ryan}},
  \bibinfo {author} {\bibfnamefont {M.}~\bibnamefont {Ditty}}, \bibinfo
  {author} {\bibfnamefont {F.}~\bibnamefont {Cyr-Racine}}, \bibinfo {author}
  {\bibfnamefont {W.}~\bibnamefont {Power}}, \bibinfo {author} {\bibfnamefont
  {N.}~\bibnamefont {Boulant}}, \bibinfo {author} {\bibfnamefont
  {T.}~\bibnamefont {Havel}}, \bibinfo {author} {\bibfnamefont {D.~G.}\
  \bibnamefont {Cory}}, \ and\ \bibinfo {author} {\bibfnamefont
  {R.}~\bibnamefont {Laflamme}},\ }\href {\doibase
  10.1103/PhysRevLett.96.170501} {\bibfield  {journal} {\bibinfo  {journal}
  {Physical Review Letters}\ }\textbf {\bibinfo {volume} {96}},\ \bibinfo
  {pages} {170501} (\bibinfo {year} {2006})}\BibitemShut {NoStop}%
\bibitem [{\citenamefont {Cappellaro}(2014)}]{cappellaro_implementation_2014}%
  \BibitemOpen
  \bibfield  {author} {\bibinfo {author} {\bibfnamefont {P.}~\bibnamefont
  {Cappellaro}},\ }in\ \href
  {http://link.springer.com/chapter/10.1007/978-3-642-39937-4_6} {\emph
  {\bibinfo {booktitle} {Quantum State Transfer and Network Engineering}}},\
  \bibinfo {series and number} {Quantum Science and Technology},\ \bibinfo
  {editor} {edited by\ \bibinfo {editor} {\bibfnamefont {G.~M.}\ \bibnamefont
  {Nikolopoulos}}\ and\ \bibinfo {editor} {\bibfnamefont {I.}~\bibnamefont
  {Jex}}}\ (\bibinfo  {publisher} {Springer Berlin Heidelberg},\ \bibinfo
  {year} {2014})\ pp.\ \bibinfo {pages} {183--222}\BibitemShut {NoStop}%
\bibitem [{\citenamefont {Rao}\ \emph {et~al.}(2013)\citenamefont {Rao},
  \citenamefont {Mahesh},\ and\ \citenamefont {Kumar}}]{rao_simulation_2013}%
  \BibitemOpen
  \bibfield  {author} {\bibinfo {author} {\bibfnamefont {K.~R.~K.}\
  \bibnamefont {Rao}}, \bibinfo {author} {\bibfnamefont {T.~S.}\ \bibnamefont
  {Mahesh}}, \ and\ \bibinfo {author} {\bibfnamefont {A.}~\bibnamefont
  {Kumar}},\ }\href {http://arxiv.org/abs/1307.5220} {\bibfield  {journal}
  {\bibinfo  {journal} {{arXiv}:1307.5220 [quant-ph]}\ } (\bibinfo {year}
  {2013})}\BibitemShut {NoStop}%
\bibitem [{\citenamefont {Zhang}\ \emph {et~al.}(2009)\citenamefont {Zhang},
  \citenamefont {Cappellaro}, \citenamefont {Antler}, \citenamefont {Pepper},
  \citenamefont {Cory}, \citenamefont {Dobrovitski}, \citenamefont
  {Ramanathan},\ and\ \citenamefont {Viola}}]{zhang_nmr_2009}%
  \BibitemOpen
  \bibfield  {author} {\bibinfo {author} {\bibfnamefont {W.}~\bibnamefont
  {Zhang}}, \bibinfo {author} {\bibfnamefont {P.}~\bibnamefont {Cappellaro}},
  \bibinfo {author} {\bibfnamefont {N.}~\bibnamefont {Antler}}, \bibinfo
  {author} {\bibfnamefont {B.}~\bibnamefont {Pepper}}, \bibinfo {author}
  {\bibfnamefont {D.~G.}\ \bibnamefont {Cory}}, \bibinfo {author}
  {\bibfnamefont {V.~V.}\ \bibnamefont {Dobrovitski}}, \bibinfo {author}
  {\bibfnamefont {C.}~\bibnamefont {Ramanathan}}, \ and\ \bibinfo {author}
  {\bibfnamefont {L.}~\bibnamefont {Viola}},\ }\href {\doibase
  10.1103/PhysRevA.80.052323} {\bibfield  {journal} {\bibinfo  {journal}
  {Physical Review A}\ }\textbf {\bibinfo {volume} {80}},\ \bibinfo {pages}
  {052323} (\bibinfo {year} {2009})}\BibitemShut {NoStop}%
\bibitem [{\citenamefont {Ajoy}\ and\ \citenamefont
  {Cappellaro}(2013)}]{ajoy_quantum_2013}%
  \BibitemOpen
  \bibfield  {author} {\bibinfo {author} {\bibfnamefont {A.}~\bibnamefont
  {Ajoy}}\ and\ \bibinfo {author} {\bibfnamefont {P.}~\bibnamefont
  {Cappellaro}},\ }\href {\doibase 10.1103/PhysRevLett.110.220503} {\bibfield
  {journal} {\bibinfo  {journal} {Physical Review Letters}\ }\textbf {\bibinfo
  {volume} {110}},\ \bibinfo {pages} {220503} (\bibinfo {year}
  {2013})}\BibitemShut {NoStop}%
\bibitem [{\citenamefont {Cory}\ \emph {et~al.}(1997)\citenamefont {Cory},
  \citenamefont {Fahmy},\ and\ \citenamefont {Havel}}]{cory_ensemble_1997}%
  \BibitemOpen
  \bibfield  {author} {\bibinfo {author} {\bibfnamefont {D.~G.}\ \bibnamefont
  {Cory}}, \bibinfo {author} {\bibfnamefont {A.~F.}\ \bibnamefont {Fahmy}}, \
  and\ \bibinfo {author} {\bibfnamefont {T.~F.}\ \bibnamefont {Havel}},\ }\href
  {http://www.pnas.org/content/94/5/1634} {\bibfield  {journal} {\bibinfo
  {journal} {Proceedings of the National Academy of Sciences}\ }\textbf
  {\bibinfo {volume} {94}},\ \bibinfo {pages} {1634} (\bibinfo {year}
  {1997})}\BibitemShut {NoStop}%
\bibitem [{\citenamefont {Gershenfeld}\ and\ \citenamefont
  {Chuang}(1997)}]{gershenfeld_bulk_1997}%
  \BibitemOpen
  \bibfield  {author} {\bibinfo {author} {\bibfnamefont {N.~A.}\ \bibnamefont
  {Gershenfeld}}\ and\ \bibinfo {author} {\bibfnamefont {I.~L.}\ \bibnamefont
  {Chuang}},\ }\href {\doibase 10.1126/science.275.5298.350} {\bibfield
  {journal} {\bibinfo  {journal} {Science}\ }\textbf {\bibinfo {volume}
  {275}},\ \bibinfo {pages} {350} (\bibinfo {year} {1997})}\BibitemShut
  {NoStop}%
\bibitem [{\citenamefont {Knill}\ and\ \citenamefont
  {Laflamme}(1997)}]{knill_theory_1997}%
  \BibitemOpen
  \bibfield  {author} {\bibinfo {author} {\bibfnamefont {E.}~\bibnamefont
  {Knill}}\ and\ \bibinfo {author} {\bibfnamefont {R.}~\bibnamefont
  {Laflamme}},\ }\href {\doibase 10.1103/PhysRevA.55.900} {\bibfield  {journal}
  {\bibinfo  {journal} {Phys. Rev. A}\ }\textbf {\bibinfo {volume} {55}},\
  \bibinfo {pages} {900–911} (\bibinfo {year} {1997})}\BibitemShut {NoStop}%
\bibitem [{\citenamefont {Kaur}\ and\ \citenamefont
  {Cappellaro}(2012)}]{kaur_initialization_2012}%
  \BibitemOpen
  \bibfield  {author} {\bibinfo {author} {\bibfnamefont {G.}~\bibnamefont
  {Kaur}}\ and\ \bibinfo {author} {\bibfnamefont {P.}~\bibnamefont
  {Cappellaro}},\ }\href {\doibase 10.1088/1367-2630/14/8/083005} {\bibfield
  {journal} {\bibinfo  {journal} {New Journal of Physics}\ }\textbf {\bibinfo
  {volume} {14}},\ \bibinfo {pages} {083005} (\bibinfo {year}
  {2012})}\BibitemShut {NoStop}%
\bibitem [{\citenamefont {Mc~Hugh}\ and\ \citenamefont
  {Twamley}(2005)}]{mc_hugh_quantum_2005}%
  \BibitemOpen
  \bibfield  {author} {\bibinfo {author} {\bibfnamefont {D.}~\bibnamefont
  {Mc~Hugh}}\ and\ \bibinfo {author} {\bibfnamefont {J.}~\bibnamefont
  {Twamley}},\ }\href {\doibase 10.1103/PhysRevA.71.012315} {\bibfield
  {journal} {\bibinfo  {journal} {Physical Review A}\ }\textbf {\bibinfo
  {volume} {71}},\ \bibinfo {pages} {012315} (\bibinfo {year}
  {2005})}\BibitemShut {NoStop}%
\bibitem [{\citenamefont {Johanning}\ \emph {et~al.}(2009)\citenamefont
  {Johanning}, \citenamefont {Varón},\ and\ \citenamefont
  {Wunderlich}}]{johanning_quantum_2009}%
  \BibitemOpen
  \bibfield  {author} {\bibinfo {author} {\bibfnamefont {M.}~\bibnamefont
  {Johanning}}, \bibinfo {author} {\bibfnamefont {A.~F.}\ \bibnamefont
  {Varón}}, \ and\ \bibinfo {author} {\bibfnamefont {C.}~\bibnamefont
  {Wunderlich}},\ }\href@noop {} {\bibfield  {journal} {\bibinfo  {journal}
  {Journal of Physics B: Atomic, Molecular and Optical Physics}\ }\textbf
  {\bibinfo {volume} {42}},\ \bibinfo {pages} {154009} (\bibinfo {year}
  {2009})}\BibitemShut {NoStop}%
\bibitem [{\citenamefont {Khromova}\ \emph {et~al.}(2012)\citenamefont
  {Khromova}, \citenamefont {Piltz}, \citenamefont {Scharfenberger},
  \citenamefont {Gloger}, \citenamefont {Johanning}, \citenamefont {Varón},\
  and\ \citenamefont {Wunderlich}}]{khromova_designer_2012}%
  \BibitemOpen
  \bibfield  {author} {\bibinfo {author} {\bibfnamefont {A.}~\bibnamefont
  {Khromova}}, \bibinfo {author} {\bibfnamefont {C.}~\bibnamefont {Piltz}},
  \bibinfo {author} {\bibfnamefont {B.}~\bibnamefont {Scharfenberger}},
  \bibinfo {author} {\bibfnamefont {T.~F.}\ \bibnamefont {Gloger}}, \bibinfo
  {author} {\bibfnamefont {M.}~\bibnamefont {Johanning}}, \bibinfo {author}
  {\bibfnamefont {A.~F.}\ \bibnamefont {Varón}}, \ and\ \bibinfo {author}
  {\bibfnamefont {C.}~\bibnamefont {Wunderlich}},\ }\href {\doibase
  10.1103/PhysRevLett.108.220502} {\bibfield  {journal} {\bibinfo  {journal}
  {Physical Review Letters}\ }\textbf {\bibinfo {volume} {108}},\ \bibinfo
  {pages} {220502} (\bibinfo {year} {2012})}\BibitemShut {NoStop}%
\bibitem [{\citenamefont {Wunderlich}(2002)}]{wunderlich_conditional_2002}%
  \BibitemOpen
  \bibfield  {author} {\bibinfo {author} {\bibfnamefont {C.}~\bibnamefont
  {Wunderlich}},\ }in\ \href
  {http://link.springer.com/chapter/10.1007/978-3-662-04897-9_25} {\emph
  {\bibinfo {booktitle} {Laser Physics at the Limits}}},\ \bibinfo {editor}
  {edited by\ \bibinfo {editor} {\bibfnamefont {D.~H.}\ \bibnamefont {Figger}},
  \bibinfo {editor} {\bibfnamefont {P.~D.~C.}\ \bibnamefont {Zimmermann}}, \
  and\ \bibinfo {editor} {\bibfnamefont {P.~D.~D.}\ \bibnamefont {Meschede}}}\
  (\bibinfo  {publisher} {Springer Berlin Heidelberg},\ \bibinfo {year}
  {2002})\ pp.\ \bibinfo {pages} {261--273}\BibitemShut {NoStop}%
\bibitem [{\citenamefont {Wunderlich}\ \emph {et~al.}(2009)\citenamefont
  {Wunderlich}, \citenamefont {Wunderlich}, \citenamefont {Singer},\ and\
  \citenamefont {Schmidt-Kaler}}]{wunderlich_two-dimensional_2009}%
  \BibitemOpen
  \bibfield  {author} {\bibinfo {author} {\bibfnamefont {H.}~\bibnamefont
  {Wunderlich}}, \bibinfo {author} {\bibfnamefont {C.}~\bibnamefont
  {Wunderlich}}, \bibinfo {author} {\bibfnamefont {K.}~\bibnamefont {Singer}},
  \ and\ \bibinfo {author} {\bibfnamefont {F.}~\bibnamefont {Schmidt-Kaler}},\
  }\href {\doibase 10.1103/PhysRevA.79.052324} {\bibfield  {journal} {\bibinfo
  {journal} {Physical Review A}\ }\textbf {\bibinfo {volume} {79}},\ \bibinfo
  {pages} {052324} (\bibinfo {year} {2009})}\BibitemShut {NoStop}%
\bibitem [{\citenamefont {Porras}\ and\ \citenamefont
  {Cirac}(2004)}]{porras_effective_2004}%
  \BibitemOpen
  \bibfield  {author} {\bibinfo {author} {\bibfnamefont {D.}~\bibnamefont
  {Porras}}\ and\ \bibinfo {author} {\bibfnamefont {J.~I.}\ \bibnamefont
  {Cirac}},\ }\href {\doibase 10.1103/PhysRevLett.92.207901} {\bibfield
  {journal} {\bibinfo  {journal} {Physical Review Letters}\ }\textbf {\bibinfo
  {volume} {92}},\ \bibinfo {pages} {207901} (\bibinfo {year}
  {2004})}\BibitemShut {NoStop}%
\bibitem [{\citenamefont {Piltz}\ \emph {et~al.}(2013)\citenamefont {Piltz},
  \citenamefont {Scharfenberger}, \citenamefont {Khromova}, \citenamefont
  {Varón},\ and\ \citenamefont {Wunderlich}}]{piltz_protecting_2013}%
  \BibitemOpen
  \bibfield  {author} {\bibinfo {author} {\bibfnamefont {C.}~\bibnamefont
  {Piltz}}, \bibinfo {author} {\bibfnamefont {B.}~\bibnamefont
  {Scharfenberger}}, \bibinfo {author} {\bibfnamefont {A.}~\bibnamefont
  {Khromova}}, \bibinfo {author} {\bibfnamefont {A.~F.}\ \bibnamefont
  {Varón}}, \ and\ \bibinfo {author} {\bibfnamefont {C.}~\bibnamefont
  {Wunderlich}},\ }\href {\doibase 10.1103/PhysRevLett.110.200501} {\bibfield
  {journal} {\bibinfo  {journal} {Physical Review Letters}\ }\textbf {\bibinfo
  {volume} {110}},\ \bibinfo {pages} {200501} (\bibinfo {year}
  {2013})}\BibitemShut {NoStop}%
\bibitem [{\citenamefont {Kim}\ \emph {et~al.}(2009)\citenamefont {Kim},
  \citenamefont {Chang}, \citenamefont {Islam}, \citenamefont {Korenblit},
  \citenamefont {Duan},\ and\ \citenamefont {Monroe}}]{kim_entanglement_2009}%
  \BibitemOpen
  \bibfield  {author} {\bibinfo {author} {\bibfnamefont {K.}~\bibnamefont
  {Kim}}, \bibinfo {author} {\bibfnamefont {M.-S.}\ \bibnamefont {Chang}},
  \bibinfo {author} {\bibfnamefont {R.}~\bibnamefont {Islam}}, \bibinfo
  {author} {\bibfnamefont {S.}~\bibnamefont {Korenblit}}, \bibinfo {author}
  {\bibfnamefont {L.-M.}\ \bibnamefont {Duan}}, \ and\ \bibinfo {author}
  {\bibfnamefont {C.}~\bibnamefont {Monroe}},\ }\href {\doibase
  10.1103/PhysRevLett.103.120502} {\bibfield  {journal} {\bibinfo  {journal}
  {Physical Review Letters}\ }\textbf {\bibinfo {volume} {103}},\ \bibinfo
  {pages} {120502} (\bibinfo {year} {2009})}\BibitemShut {NoStop}%
\bibitem [{\citenamefont {Lubasch}\ \emph {et~al.}(2011)\citenamefont
  {Lubasch}, \citenamefont {Murg}, \citenamefont {Schneider}, \citenamefont
  {Cirac},\ and\ \citenamefont {Bañuls}}]{lubasch_adiabatic_2011}%
  \BibitemOpen
  \bibfield  {author} {\bibinfo {author} {\bibfnamefont {M.}~\bibnamefont
  {Lubasch}}, \bibinfo {author} {\bibfnamefont {V.}~\bibnamefont {Murg}},
  \bibinfo {author} {\bibfnamefont {U.}~\bibnamefont {Schneider}}, \bibinfo
  {author} {\bibfnamefont {J.~I.}\ \bibnamefont {Cirac}}, \ and\ \bibinfo
  {author} {\bibfnamefont {M.-C.}\ \bibnamefont {Bañuls}},\ }\href {\doibase
  10.1103/PhysRevLett.107.165301} {\bibfield  {journal} {\bibinfo  {journal}
  {Physical Review Letters}\ }\textbf {\bibinfo {volume} {107}},\ \bibinfo
  {pages} {165301} (\bibinfo {year} {2011})}\BibitemShut {NoStop}%
\bibitem [{\citenamefont {Trotzky}\ \emph {et~al.}(2008)\citenamefont
  {Trotzky}, \citenamefont {Cheinet}, \citenamefont {Fölling}, \citenamefont
  {Feld}, \citenamefont {Schnorrberger}, \citenamefont {Rey}, \citenamefont
  {Polkovnikov}, \citenamefont {Demler}, \citenamefont {Lukin},\ and\
  \citenamefont {Bloch}}]{trotzky_time-resolved_2008}%
  \BibitemOpen
  \bibfield  {author} {\bibinfo {author} {\bibfnamefont {S.}~\bibnamefont
  {Trotzky}}, \bibinfo {author} {\bibfnamefont {P.}~\bibnamefont {Cheinet}},
  \bibinfo {author} {\bibfnamefont {S.}~\bibnamefont {Fölling}}, \bibinfo
  {author} {\bibfnamefont {M.}~\bibnamefont {Feld}}, \bibinfo {author}
  {\bibfnamefont {U.}~\bibnamefont {Schnorrberger}}, \bibinfo {author}
  {\bibfnamefont {A.~M.}\ \bibnamefont {Rey}}, \bibinfo {author} {\bibfnamefont
  {A.}~\bibnamefont {Polkovnikov}}, \bibinfo {author} {\bibfnamefont {E.~A.}\
  \bibnamefont {Demler}}, \bibinfo {author} {\bibfnamefont {M.~D.}\
  \bibnamefont {Lukin}}, \ and\ \bibinfo {author} {\bibfnamefont
  {I.}~\bibnamefont {Bloch}},\ }\href {\doibase 10.1126/science.1150841}
  {\bibfield  {journal} {\bibinfo  {journal} {Science}\ }\textbf {\bibinfo
  {volume} {319}},\ \bibinfo {pages} {295} (\bibinfo {year}
  {2008})}\BibitemShut {NoStop}%
\bibitem [{\citenamefont {Fukuhara}\ \emph
  {et~al.}(2013{\natexlab{b}})\citenamefont {Fukuhara}, \citenamefont
  {Kantian}, \citenamefont {Endres}, \citenamefont {Cheneau}, \citenamefont
  {Schauß}, \citenamefont {Hild}, \citenamefont {Bellem}, \citenamefont
  {Schollwöck}, \citenamefont {Giamarchi}, \citenamefont {Gross},
  \citenamefont {Bloch},\ and\ \citenamefont {Kuhr}}]{fukuhara_quantum_2013}%
  \BibitemOpen
  \bibfield  {author} {\bibinfo {author} {\bibfnamefont {T.}~\bibnamefont
  {Fukuhara}}, \bibinfo {author} {\bibfnamefont {A.}~\bibnamefont {Kantian}},
  \bibinfo {author} {\bibfnamefont {M.}~\bibnamefont {Endres}}, \bibinfo
  {author} {\bibfnamefont {M.}~\bibnamefont {Cheneau}}, \bibinfo {author}
  {\bibfnamefont {P.}~\bibnamefont {Schauß}}, \bibinfo {author} {\bibfnamefont
  {S.}~\bibnamefont {Hild}}, \bibinfo {author} {\bibfnamefont {D.}~\bibnamefont
  {Bellem}}, \bibinfo {author} {\bibfnamefont {U.}~\bibnamefont {Schollwöck}},
  \bibinfo {author} {\bibfnamefont {T.}~\bibnamefont {Giamarchi}}, \bibinfo
  {author} {\bibfnamefont {C.}~\bibnamefont {Gross}}, \bibinfo {author}
  {\bibfnamefont {I.}~\bibnamefont {Bloch}}, \ and\ \bibinfo {author}
  {\bibfnamefont {S.}~\bibnamefont {Kuhr}},\ }\href {\doibase
  10.1038/nphys2561} {\bibfield  {journal} {\bibinfo  {journal} {Nature
  Physics}\ }\textbf {\bibinfo {volume} {9}},\ \bibinfo {pages} {235} (\bibinfo
  {year} {2013}{\natexlab{b}})}\BibitemShut {NoStop}%
\bibitem [{\citenamefont {Clark}\ \emph {et~al.}(2005)\citenamefont {Clark},
  \citenamefont {Moura~Alves},\ and\ \citenamefont
  {Jaksch}}]{clark_efficient_2005}%
  \BibitemOpen
  \bibfield  {author} {\bibinfo {author} {\bibfnamefont {S.~R.}\ \bibnamefont
  {Clark}}, \bibinfo {author} {\bibfnamefont {C.}~\bibnamefont {Moura~Alves}},
  \ and\ \bibinfo {author} {\bibfnamefont {D.}~\bibnamefont {Jaksch}},\
  }\href@noop {} {\bibfield  {journal} {\bibinfo  {journal} {New Journal of
  Physics}\ }\textbf {\bibinfo {volume} {7}},\ \bibinfo {pages} {124} (\bibinfo
  {year} {2005})}\BibitemShut {NoStop}%
\bibitem [{\citenamefont {Duan}\ \emph {et~al.}(2003)\citenamefont {Duan},
  \citenamefont {Demler},\ and\ \citenamefont {Lukin}}]{duan_controlling_2003}%
  \BibitemOpen
  \bibfield  {author} {\bibinfo {author} {\bibfnamefont {L.~M.}\ \bibnamefont
  {Duan}}, \bibinfo {author} {\bibfnamefont {E.}~\bibnamefont {Demler}}, \ and\
  \bibinfo {author} {\bibfnamefont {M.~D.}\ \bibnamefont {Lukin}},\ }\href@noop
  {} {\bibfield  {journal} {\bibinfo  {journal} {Physical review letters}\
  }\textbf {\bibinfo {volume} {91}},\ \bibinfo {pages} {90402} (\bibinfo {year}
  {2003})}\BibitemShut {NoStop}%
\bibitem [{\citenamefont {Biedenharn}\ and\ \citenamefont
  {Louck}(1981)}]{biedenharn_angular_1981}%
  \BibitemOpen
  \bibfield  {author} {\bibinfo {author} {\bibfnamefont {L.}~\bibnamefont
  {Biedenharn}}\ and\ \bibinfo {author} {\bibfnamefont {J.~D.}\ \bibnamefont
  {Louck}},\ }\href@noop {} {\emph {\bibinfo {title} {Angular Momentum in
  Quantum Physics: Theory and Application, Encyclopedia of Mathematics and its
  Applications}}}\ (\bibinfo  {publisher} {Addison-Wesley, Englewood Cliffs},\
  \bibinfo {year} {1981})\BibitemShut {NoStop}%
\bibitem [{\citenamefont {Banchi}\ and\ \citenamefont
  {Vaia}(2013)}]{banchi_spectral_2013}%
  \BibitemOpen
  \bibfield  {author} {\bibinfo {author} {\bibfnamefont {L.}~\bibnamefont
  {Banchi}}\ and\ \bibinfo {author} {\bibfnamefont {R.}~\bibnamefont {Vaia}},\
  }\href@noop {} {\bibfield  {journal} {\bibinfo  {journal} {Journal of
  Mathematical Physics}\ }\textbf {\bibinfo {volume} {54}},\ \bibinfo {pages}
  {43501} (\bibinfo {year} {2013})}\BibitemShut {NoStop}%
\end{thebibliography}
\end{document}